\documentclass[aps,pra,reprint,nofootinbib,superscriptaddress,twocolumn,showpacs,showkeys,longbibliography,amsmath,amssymb]{revtex4-2}
\usepackage{graphicx}
\usepackage{dcolumn}
\usepackage{bm}
\usepackage{braket}
\usepackage{subfigure}
\usepackage[colorlinks,bookmarks=false,citecolor=blue,linkcolor=red,urlcolor=blue]{hyperref}
\usepackage[english]{babel}
\usepackage{changes}
\usepackage{amsmath}
\usepackage{dcolumn}
\usepackage{appendix}
\usepackage{float}
\usepackage{makecell}
\usepackage{multirow}
\begin{document}
\preprint{APS/123-QED}
\title{Cornwall-Jackiw-Tomboulis effective field theory to nonuniversal equation of state of an ultracold Bose gas}
\author{Yi Zhang}
\affiliation{Department of Physics, Zhejiang Normal University, Jinhua 321004, People's Republic of China}
\author{Zhaoxin Liang}\email[Corresponding author:~] {zhxliang@zjnu.edu.cn}
\affiliation{Department of Physics, Zhejiang Normal University, Jinhua 321004, People's Republic of China}
\date{\today}

\begin{abstract}
The equation of state (EOS) serves as a cornerstone in elucidating the properties of quantum many-body systems.
A recent highlight along this research line consists of the derivation of the nonuniversal Lee-Huang-Yang (LHY) 
EOS for an ultracold quantum bosonic gas with finite-range interatomic interactions using one-loop effective path-integral field theory 
[\href{https://doi.org/10.1103/PhysRevLett.118.130402}{L. Salasnich, Phys. Rev. Lett. 118, 130402 (2017)}].
The purpose of this work is to extend Salasnich's pioneering work to uncover beyond-LHY corrections to the EOS by employing the Cornwall-Jackiw-Tomboulis (CJT) effective field theory, 
leveraging its two-loop approximation. 
In this end, we expand Salasnich's remarkable findings of EOS to the next leading order characterized by $\left(\rho a_{\text{s}}^{3}\right)^{2}$, with $\rho$ and $a_{\text{s}}$ being the density and the $s$-wave scattering length.  
Notably, we derive analytical expressions for quantum depletion and chemical potential, 
representing the next-to-LHY corrections to nonuniversal EOS induced by finite-range effects.
Moreover, we propose an experimental protocol of observing the nonuniversal next-to-LHY corrections to the EOS by
calculating fractional frequency shifts in the breathing modes. 
The nonuniversal beyond-LHY EOS in this work paves the way of using LHY effects in quantum simulation experiments and for investigations beyond the LHY regime.
\end{abstract}
\maketitle

\section{Introduction}\label{Introduction}
The equation of state (EOS) of quantum physical systems, arising from quantum fluctuations, occupies a pivotal position at the very core of quantum many-body physics~\cite{Nagaosa1999,Pitaevskii2016,Peskin1995}.
A prototypical illustration lies within the realm of quantum droplets~\cite{Bulgac2002,Petrov2015,Tanzi2019} in the context of ultracold quantum gases.
The basic principle underlying the stabilization of quantum droplets hinges on a delicate equilibrium between the attractive mean-field force and the repulsive force emanating from quantum fluctuations as described by the EOS~\cite{Igor2016,L.Chomaz2016,Wenzel2017,G.Semeghini2018,Cabrera2018,D'Errico2019}.
Consequently, delving into the EOS can provide invaluable insights into the fundamental properties of quantum droplets, as evidenced across diverse systems, spanning from dipolar Bose gases~\cite{Igor2016,L.Chomaz2016,Wenzel2017} to Bose-Bose mixtures~\cite{G.Semeghini2018,Cabrera2018,D'Errico2019}.

The interest in the EOS of ultracold quantum gases, particularly underscored by the aforementioned quantum droplets, can be traced back to the seminal works~\cite{Lee1957,Yang1957} in the 1950s. Particularly, the Lee-Huang-Yang (LHY) correction to the EOS of weakly-interacting bosonic systems was first explored in Ref.~\cite{Yang1957}.
In more details, the chemical potential has been derived to be proportional to $[32/3\sqrt{\pi}]\left(\rho a_{\text{s}}^{3}\right)^{1/2}$ with $a_{\text{s}}$~\cite{Dalfovo1999,Roth2001} and $\rho$ being the $s$-wave scattering length and the atomic density, respectively.
Furthermore, the next-order correction to the energy density incorporates the universal quantum effect stemming from three-body correlations, yielding a term of $\frac{8}{3}\left(4\pi-3\sqrt{3}\right)\rho a_{\text{s}}^{3}\ln\left(\rho  a_{\text{s}}^{3}\right)$, as outlined in Ref.~\cite{Wu1959}.
Up to now, the highest-order term in the EOS has been derived in Ref.~\cite{Van2022}, which is given by  $[-512/3\pi]\left(\rho  a_{\text{s}}^{3}\right)+[8192/9\pi^{3/2}]\left(\rho  a_{\text{s}}^{3}\right)^{3/2}$. At present, there are the endless and ongoing research interests and efforts in obtaining beyond the mean-field terms of EOS of weakly-interacting bosonic systems motivated by the advent of well-controlled mixtures of quantum gases with tunable interaction strengths~\cite{Mordini2020,Skov2021,Lavoine2021,Cominotti2023,Science2022}.

The EOS obtained in Refs.~\cite{Lee1957,Yang1957,Dalfovo1999,Roth2001,Wu1959,Van2022} are characterized by the universality, i.e. in more details, a single parameter $a_{\text{s}}$, eloquently encapsulates both the intricacies of the two-body interaction and, by extension, the overarching physics governing the many-body systems \cite{Wilhelm2008}.
In sharp contrast, the nonuniversal EOS depends on other than the parameter $a_{\text{s}}$. 
The EOS of Bose gases becomes nonuniversal when the finite-range effects of the interatomic potential~\cite{Braaten2001,Cappellaro2017,Lorenzi2023,Ye2024,Yutao2024} is taken into account.
In Refs.~\cite{Cappellaro2017,Tononi2018}, the analytical expressions of nonuniversal EOS are derived as $[-64\sqrt{\pi}r_{\text{s}}/ a_{\text{s}}]\rho\left(\rho  a_{\text{s}}^{3}\right)^{3/2}$, featuring a nonuniversal LHY term in the quantum depletion.
Given the tunability of the scattering length $a_\text{s}$~\cite{Cornish_2000,Pethick_2002} through magnetic and optical Feshbach resonances~\cite{Chin2010,HaibinWU2012,Wu2012} in ultracold atomic gases, the nonuniversal consequences stemming from the finite-range parameter $r_\text{s}$ are of paramount importance.
Consequently, an immediate challenge is referred as to calculating the next-to-LHY-order correction for the nonuniversal EOS.

The second impetus behind this paper stems from the measurement of collective excitation frequencies, which has emerged as an indispensable and highly accurate instrument for delving into the intricate EOS of atomic Bose-Einstein condensates (BECs) with unprecedented precision~\cite{Dalfovo1999,Morsch2006,Mordini2020,Skov2021,Lavoine2021,Cominotti2023,Science2022}.
This approach not only solidifies the validity of mean-field predictions but also stands as an exceedingly potent technique to explore effects transcending the mean-field paradigm~\cite{Stringari1996,Pitaevskii1998}.
From a theoretical standpoint, a gaseous BECs system can be aptly modeled by a single macroscopic wave function, thereby facilitating the derivation of clear-cut hydrodynamic formulations that yield analytical or semi-analytical insights into the dynamical characteristics of BECs systems. Thus, a timely and natural question arises as to observe the frequency shifts in the collective excitations induced by the nonuniversal beyond-LHY EOS, although tuning the finite-range interatomic interactions in this case remains experimentally challenging.

In this work, using Cornwall-Jackiw-Tomboulis (CJT) effective field theory~\cite{Cornwall1974,Amelino1993,Teresi2013,Sharma2022}, we are interested in the nonuniversal EOS~\cite{Salasnich2017,Cappellaro2017,Tononi2018,Ye2024,Yutao2024} of a three-dimensional (3D) Bose gas with finite-range effective interactions at absolute zero temperature. Accordingly, we derive the analytical expressions of the quantum depletion and the chemical potential up to the order of $\left(\rho a_{\text{s}}^{3}\right)^{2}$.
Our results not only include Salasnich’s remarkable EOS, but also give rise to nonuniversal beyond-LHY terms.
Moreover, we explore the physical consequences of the nonuniversal beyond-LHY terms in EOS. As such, 
we extend the superfluid hydrodynamic equations by incorporating the aforementioned next-to-LHY corrections to the nonuniversal EOS.
Leveraging these refined hydrodynamic equations, we delve into the fractional frequency
shifts in the breathing modes, which are observable within the current experimental facilities.
Observing this frequency shifts induced by the nonuniversal beyond-LHY EOS paves the way for a deeper understanding of quantum fluctuation of quantum many-body systems.

The paper is structured as follows.
In Sec.~\ref{MSAH}, we revisit the key principles of the CJT effective field theory and subsequently derive the corresponding effective potential for the model system.
In Sec.~\ref{NEST}, we utilize the CJT effective potential within a two-loop approximation to deduce analytical expressions for the nonuniversal EOS.
Sec.~\ref{CETN} presents the fractional shift in the breathing mode frequency, aided by the next-to-LHY corrections to the nonuniversal chemical potential.
Finally, in Sec.~\ref{DAC}, we provide a comprehensive summary of our paper and discuss the potential experimental conditions for realizing our proposed scenario.

\section{Cornwall-Jackiw-Tomboulis effective field theory}\label{MSAH}
In this work, we are interested in a 3D weakly-interacting Bose gas, paying particular attention to the finite-range effects of the interatomic potential~\cite{Lorenzi2023}. To investigate this system, we employ the path-integral formalism~\cite{Nagaosa1999,Pitaevskii2016}, and the corresponding Euclidean partition function of the model system is presented below
\begin{equation}
\label{partition function}
\mathcal{Z}=\int\mathcal{D}\left[\mathbf{\Phi},\mathbf{\Phi}^{*}\right]\exp\left\{ -\frac{S\left[\mathbf{\Phi},\mathbf{\Phi}^{*}\right]}{\hbar}\right\}, 
\end{equation}
with the action functional $S\left[\mathbf{\Phi},\mathbf{\Phi}^{*}\right]=\int_{0}^{\beta\hbar}d\tau\int d^{3}\mathbf{r}\mathcal{L}$ in Eq.~(\ref{partition function}).
Here, the concrete Lagrangian density $\mathcal{L}$ denotes as follows \cite{Salasnich2017,Cappellaro2017,Tononi2018,SALASNICH_2016}
\begin{widetext}
\begin{eqnarray}
\label{Lagrangian density}
\mathcal{L}\left[\mathbf{\Phi},\mathbf{\Phi}^{*}\right]
=\mathbf{\Phi}^{*}\left(\mathbf{r},\tau\right)\left[\hbar\partial_{\tau}-\frac{\hbar^{2}\nabla^{2}}{2m}-\mu\right]\mathbf{\Phi}\left(\mathbf{r},\tau\right)+\frac{g_{0}}{2}\left|\mathbf{\Phi}\left(\mathbf{r},\tau\right)\right|^{4}-\frac{g_{2}}{2}\left|\mathbf{\Phi}\left(\mathbf{r},\tau\right)\right|^{2}\nabla^{2}\left|\mathbf{\Phi}\left(\mathbf{r},\tau\right)\right|^{2}. 
\end{eqnarray}
\end{widetext}
In Equation~(\ref{Lagrangian density}), the complex field $\mathbf{\Phi}\left(\mathbf{r},\tau\right)$ represents the atomic bosons, varying in both space $\mathbf{r}$ and imaginary time $\tau$. Here, $\mu$ signifies the chemical potential, while $\beta=1/k_{\text{B}}T$ defines 
the inverse of the thermal energy scale, with $k_{\text{B}}$ representing the Boltzmann constant
and $T$ denoting the temperature of the BECs. The parameters $g_{0}=4\pi\hbar^{2} a_{\text{s}}/m$ and $g_{2}=2\pi\hbar^{2} a_{\text{s}}^{2}r_{\text{s}}/m$ \cite{Salasnich2017}, with $a_{\text{s}}$ and $r_\text{s}$ being the $s$-wave scattering length and the finite-range length respectively.

Note that the LHY correction to nonuniversal EOS of Lagrangian density functional (\ref{Lagrangian density}) has already been derived \cite{Cappellaro2017,Tononi2018} within one-loop approximation, i.e. rewriting the partition function (\ref{partition function}) in the form of $\mathcal{Z}=e^{-\beta \mathcal{V}V_{\text{eff}}[\phi]}$, with $\mathcal{V}$ being the volume of the system.
In this context, the effective potential $V_{\text{eff}}[\phi]$ \cite{HAUGSET_1998} within one-loop approximation solely relies on $\phi(x)$, which is the expected value of the quantum field $\hat{\mathbf{\Phi}}(x)$. 

In contrast, the emphasis and value of this paper lie in employing the CJT \cite{Amelino1993,Sharma2022,Teresi2013,PHAT_2009,HAUGSET_1998} effective action approach or the two particle-irreducible framework, surpassing the limitations of the one-loop approximation~\cite{HAUGSET_1998}. 
CJT theory enables us to compute the beyond-LHY corrections to the nonuniversal EOS for the Lagrangian density functional (\ref{Lagrangian density}). 
A comprehensive introduction to the CJT effective field theory is provided in Appendix \ref{CEAA}.

The central step of CJT effective field theory~\cite{Cornwall1974,Amelino1993,Teresi2013,Sharma2022} is to obtain the effective potential denoted as $V_{\text{eff}}[\phi,G]$ (see Eq.~(\ref{V_CJT}) in Appendix \ref{CEAA}) meticulously by taking both the background field $\phi(x)$ and the dressed propagator $G(x,y)$ into account. Here, $G(x,y)$ is a potential expectation value of the time-ordered product $T\hat{\mathbf{\Phi}}^{\dagger}(x)\hat{\mathbf{\Phi}}(y)$. Then, physical solutions are ascertained by ensuring that the generalized effective potential satisfies the stationarity conditions: $\delta V_{\text{eff}}/\delta \phi(x)=0$, $\delta V_{\text{eff}}/\delta G(x,y)=0$. Finally, at the core of the CJT effective field theory lies in the self-consistent loop expansion of the effective potential $V_{\text{eff}}$, intricately tied to the full propagator $G$.
This expansion offers a powerful tool for systematically exploring higher-order corrections, thereby enhancing the accuracy and predictive capabilities of the system with finite-range effects taken into account.
We remark that our work, together with Refs. \cite{Cappellaro2017,Tononi2018}, provides a reasonable description of the nonuniversal EOS of a 3D interacting Bose gas with finite-range effects of the interatomic potential.

The subsequent goal of Sec.~\ref{MSAH} is to derive the effective potential $V_{\text{eff}}$ for functional (\ref{Lagrangian density}) within the framework of CJT effective field theory. Then, the obtained effective potential $V_{\text{eff}}$ is used to calculate the nonuniversal beyond-LHY EOS. 

The starting point of the CJT effective field theory commences with expressing the field $\mathbf{\Phi}$ of functional (\ref{Lagrangian density}) as a superposition of the condensate field $\phi_{0}$ and real fluctuation fields of $\phi_{1}$ and $\phi_{2}$, i.e. 
$\mathbf{\Phi}=\left(\phi_{0}+\phi_{1}+i\phi_{2}\right)/\sqrt{2}$.
A rigorous methodology to attain $V_{\text{eff}}$ involves executing a double Legendre transformation on the action functional $S\left[\mathbf{\Phi},\mathbf{\Phi}^{*}\right]$, subjecting it to the conditions $\delta V_{\text{eff}}/\delta\phi_{0}=0$ and $\delta V_{\text{eff}}/\delta G=0$,
where $G$ represents a relevant variable (such as the two-point function or propagator) that may need to be optimized simultaneously with $\phi_{0}$.
This ensures the effective potential accurately captures the dynamics of the system, incorporating both condensate and fluctuation effects.

Adhering closely to the established CJT effective field theory as outlined in Refs.~\cite{Phat_2007,PHat_2008,PHAT_2009,Song2022,Van2022},
we proceed by performing Fourier transformations on the fluctuation fields $\phi_1$ and $\phi_2$
to map them into the momentum-frequency domain.
Subsequently, we select the Luttinger-Ward functional~\cite{HAUGSET_1998}, denoted as $\Phi\left[\phi_0,G\right]$, as the foundational element.
Consequently, the effective potential $V_{\text{eff}}$ corresponding to partition function
in Eq.~(\ref{partition function}) can be analytically formulated as
(see the detailed derivation in Eq.~(\ref{the orginal Veff}) in Appendix~\ref{DDTE})
\begin{eqnarray}
\label{Veff_phi_0,P_11,P_22}
V_{\text{eff}}\left[\phi_{0},G\right]= & -&\frac{\mu}{2}\phi_{0}^{2}+\frac{g_{0}}{8}\phi_{0}^{4}\nonumber \\
 & +&\frac{1}{2}\int_{\beta}\text{Tr}\left[\ln G^{-1}\left(k\right)+G_{0}^{-1}\left(k\right)G\left(k\right)-\mathbf{1}\right]\nonumber \\
 & +&\frac{3g_{0}}{8}\left(P_{11}^{2}+P_{22}^{2}\right)+\frac{g_{0}}{4}P_{11}P_{22}.
\end{eqnarray}
In Equation (\ref{Veff_phi_0,P_11,P_22}), $k$ denotes the magnitude of the wave vector $\mathbf{k}$.
The quantities $G^{-1}_{0}\left(k\right)$ and $G^{-1}\left(k\right)$ represent the inverse propagators within the one-loop and two-loop approximations, respectively.
The notation $\int_{\beta}$ encapsulates the integration over momentum space combined with a summation over bosonic Matsubara frequencies, specifically given by $\int_{\beta}=\frac{1}{\beta}\sum_{n=-\infty}^{+\infty}\int\frac{d\mathbf{k}}{\left(2\pi\right)^{3}}$;
$\omega_{n}=2\pi n\beta^{-1}$ are the Matsubara frequencies. 
The functions $P_{aa}=\int_{\beta}G_{aa}\left(k\right)$ are termed momentum integrals, with $a$ indexing the two fields $\left(1,2\right)$.

Minimizing the CJT effective potential $V_{\text{eff}}$ in Eq.~(\ref{Veff_phi_0,P_11,P_22}) with respect to the components of the propagator $G\left(k\right)$, we obtain
\begin{equation}
\label{propergater G-1_not in appendix}
G^{-1}\left(k\right)=G_{0}^{-1}\left(k\right)+\Sigma.
\end{equation}
In Equation~(\ref{propergater G-1_not in appendix}), the context form of $G_{0}^{-1}$ can be written as (see the detailed derivation in Eq.~(\ref{G0,-1}) in Appendix~\ref{oneloop})
 \begin{equation}
 \label{G0,-1 not in appendix}
\!\!G_{0}^{-1}\!\left(k\right)  \!=\!\begin{bmatrix}\!\frac{\hbar^{2}k^{2}}{2m}\!-\!\mu\!+\!\frac{3g_{0}}{2}\phi_{0}^{2}\!+g_{2}\phi_{0}^{2}k^{2} & \!\!-\omega_{n}\\
\!\!\omega_{n} & \frac{\hbar^{2}k^{2}}{2m}\!-\!\mu\!+\!\frac{g_{0}}{2}\phi_{0}^{2}
\end{bmatrix}.\!\!
\end{equation}  
Meanwhile, $\Sigma$ in Eq.~(\ref{propergater G-1_not in appendix}) represents the self-energy matrix in the form of
\begin{equation}
\label{sum not in appendix}
\Sigma=\begin{bmatrix}\Sigma_{1} & 0\\
0 & \Sigma_{2}
\end{bmatrix},
\end{equation}
with the matrix entries $\Sigma_{1}$ and $\Sigma_{2}$ reading
\begin{subequations}
\begin{eqnarray}
\Sigma_{1}&=&\frac{3g_{0}}{2}P_{11}+\frac{g_{0}}{2}P_{22},\label{sum1 not in appendix}\\
\Sigma_{2}&=&\frac{3g_{0}}{2}P_{22}+\frac{g_{0}}{2}P_{11}.\label{sum2 not in appendix}
\end{eqnarray}
\end{subequations} 

Before proceeding with further calculations based on Eq.~(\ref{Veff_phi_0,P_11,P_22}), 
we conduct a crucial verification to ensure that Eq.~(\ref{Veff_phi_0,P_11,P_22}) can indeed reproduce the previous findings presented in Refs.~\cite{Salasnich2017,Cappellaro2017,Tononi2018}.
Specifically, under the conditions where the propagator $G\left(k\right)$ reduces to its bare form $G_0\left(k\right)$ 
and both $P_{11}$ and $P_{22}$ vanish, the effective potential simplifies significantly to align with the one-loop approximation scenario. 
This simplification yields $V_{\text{eff}}\left[\phi_{0},G_0\right]=-\frac{\mu}{2}\phi_{0}^{2}+\frac{g_{0}}{8}\phi_{0}^{4}
+\frac{1}{2}\int_{\beta}\text{Tr}\ln \left[G_{0}^{-1}\left(k\right)\right]$ within the context of the path-integral formalism. Then, after minimizing $V_{\text{eff}}$ with respect to order parameter $\phi_{0}$ and applying the thermodynamic relationship 
$\rho=-\partial V_{\text{eff}}/\partial \phi_{0}$, the nonuniversal EOS~\cite{Salasnich2017,Cappellaro2017,Tononi2018,Ye2024,Yutao2024} can be obtained within one-loop approximation.
Note that the CJT effective field theory not only introduces two-particle irreducible (2PI) terms in the last line of Eq.~(\ref{Veff_phi_0,P_11,P_22}) but also modifies the propagator $G$ in a more intricate manner, 
affecting the second term of $V_{\text{eff}}$ in Eq.~(\ref{Veff_phi_0,P_11,P_22}) in a complex and non-trivial way.

\section{NONUNIVERSAL EOS: chemical potential and quantum depletion}\label{NEST}
In the preceding Sec. \ref{MSAH}, we have delineated the framework of the CJT effective field theory.
Moving forward, in Sec. \ref{NEST}, our objective is to derive the explicit analytical expressions for the next-to-LHY-order correction to the nonuniversal EOS of a 3D Bose gas, utilizing the CJT effective potential of Eq.~(\ref{Veff_phi_0,P_11,P_22}) within two-loop approximation. 
The starting point for this endeavor is to deduce the $V_{\text{eff}}$ in Eq.~(\ref{Veff_phi_0,P_11,P_22}) as (refer to Appendix~\ref{DDTE} for a more comprehensive derivation),
\begin{widetext}
\begin{equation}\label{the final Veff}
V_{\text{eff}}=V_{0}+\frac{1}{2}\int_{\beta}\text{\text{Tr}}\left[\ln G^{-1}\left(k\right)\right]
  +\frac{g_{0}}{8}\left(P_{11}^{2}+P_{22}^{2}\right)+\frac{3g_{0}}{4}P_{11}P_{22}
  +\frac{1}{2}\left(-\mu+\frac{3g_{0}}{2}\phi_{0}^{2}-M^{2}\right)P_{11}
  +\frac{1}{2}\left(-\mu+\frac{g_{0}}{2}\phi_{0}^{2}\right)P_{22}.
\end{equation}
\end{widetext}
In Equation (\ref{the final Veff}), $V_0=-\frac{\mu}{2}\phi_{0}^{2}+\frac{g_{0}}{8}\phi_{0}^{4}$, $P_{11}=\frac{\sqrt{2}m^{*3/2}M^{3}}{3\pi^{2}\hbar^{3}}\sqrt{\frac{m^*}{m}}$ and $P_{22}=\frac{-m^{*3/2}M^{3}}{3\sqrt{2}\pi^{2}\hbar^{3}}\sqrt{\frac{m}{m^*}}$
(see Eqs.~(\ref{P11 about M}) and (\ref{P22 about M}) in Appendix~\ref{DDTE} for calculation details) with
$m^{*}=m/\left(1+\frac{2mg_{2}\phi_{0}^{2}}{\hbar^{2}}\right)$. 

The pivotal parameter $M$ in Eq. (\ref{the final Veff}) holds a crucial role in determining not only $P_{11}$ and $P_{22}$, but also the other constituent terms in Eq. (\ref{the final Veff}). 
Subsequently, we embark on deriving the precise expression for $M$, as outlined below:

(i). The parameter $M$ in Eq. (\ref{the final Veff}) fulfills the Schwinger-Dyson (SD) equation
(as elaborated in detail in Eq.~(\ref{Dyson equation}) in Appendix \ref{DDTE}),
\begin{equation}\label{SD equation with P11 and P22}
-\mu+\frac{3g_{0}}{2}\phi_{0}^{2}+\frac{\sqrt{2}g_{0}m^{*3/2}M^{3}}{12\pi^{2}\hbar^{3}}\left(2\sqrt{\frac{m^{*}}{m}}-3\sqrt{\frac{m}{m^{*}}}\right)=M^{2}.
\end{equation}
Equation (\ref{SD equation with P11 and P22}) contains three variables, i.e. $M$, $\mu$, and $\phi_0$. 
To ascertain their values, we require two additional equations in conjunction with Eq. (\ref{SD equation with P11 and P22}).

(ii). Next, we proceed to seek for the second equation between $M$, $\mu$, and $\phi_0$. Specifically, the chemical potential of $\mu$ satisfies the gap equation by setting $\delta V_{\text{eff}}/\delta\phi_{0}=0$
(for a detailed exposition, refer to Eq. (\ref{the Eq. of mu}) in Appendix \ref{DDTE}),
\begin{equation}\label{mu gap equation with P11 and P22}
  -\mu+\frac{g_{0}}{2}\phi_{0}^{2}+\frac{\sqrt{2}g_{0}m^{*3/2}M^{3}}{12\pi^{2}\hbar^{3}}\left(6\sqrt{\frac{m^{*}}{m}}-\sqrt{\frac{m}{m^{*}}}\right)=0.
\end{equation} 

(iii). Finally, the third equation originated by  
determining the atomic density of $\rho$, can be obtained by taking the first-order derivative of the $V_{\text{eff}}$ in
Eq.~(\ref{the final Veff}) 
with respect to $\mu$,
\begin{equation}
\label{the thermodynamic relation of rho}
\text{\ensuremath{\rho}}=-\frac{\partial V_{\text{eff}}}{\partial\mu}=\frac{\phi_{0}^{2}}{2}+\frac{\sqrt{2}m^{*3/2}M^{3}}{12\pi^{2}\hbar^{3}}\left(2\sqrt{\frac{m^{*}}{m}}-\sqrt{\frac{m}{m^{*}}}\right).
\end{equation}
Equations~(\ref{SD equation with P11 and P22}), (\ref{mu gap equation with P11 and P22})
and (\ref{the thermodynamic relation of rho}) constitute a closed set of equations.
By eliminating the variables $\mu$ and $\phi_0$ from Eq.~(\ref{SD equation with P11 and P22}) with the aid of Eqs.~(\ref{mu gap equation with P11 and P22}) and~(\ref{the thermodynamic relation of rho}), 
we obtain a cubic equation solely in terms of $M$,
\begin{equation}
\label{the equation of $M$}
M^{3}+\frac{\left(1+4\frac{mg_{2}}{\hbar^{2}}\rho\right)^{3/2}}{8\sqrt{2}a_{\text{s}}m^{*1/2}}M^{2}-\frac{3\pi^{2}\hbar^{3}\left(1+4\frac{mg_{2}}{\hbar^{2}}\rho\right)^{1/2}}{\sqrt{2}m^{*3/2}}\rho=0,
\end{equation}
with $g_{0}=4\pi\hbar^{2}a_{\text{s}}/m$ and $m^{*}\simeq m/\left(1+4m\frac{g_{2}}{\hbar^{2}}\rho\right)$.

By solving Equation~(\ref{the equation of $M$}) using perturbation theory (details provided in Appendix \ref{DCTS}) 
and substituting $g_2$ with $2\pi\hbar^{2}a_{\text{s}}^2r_{\text{s}}/m$, we derive the analytical expression of $M$ expanded as the gas parameter $\rho a_{\text{s}}^{3}$
\begin{widetext}
\begin{equation}
\label{the solution of $M$}
M=\sqrt{2\rho g_{0}}\left\{1-\frac{16}{3\sqrt{\pi}\left(1+8\pi\frac{r_{\text{s}}}{a_{\text{s}}}\rho a_{\text{s}}^{3}\right)^{2}}\sqrt{\rho a_{\text{s}}^{3}}\left[1-\frac{40}{3\sqrt{\pi}\left(1+8\pi\frac{r_{\text{s}}}{a_{\text{s}}}\rho a_{\text{s}}^{3}\right)^{2}}\sqrt{\rho a_{\text{s}}^{3}}\right]+\mathcal{O}\left[\left(\rho a_{\text{s}}^{3}\right)^{3/2}\right]\right\}.
\end{equation}
\end{widetext}

In Equation~(\ref{the solution of $M$}), the effective mass $M$, emerges as a pivotal parameter characterizing weakly-interacting BECs. 
Our findings, encapsulated in Eq.~(\ref{the solution of $M$}) accurately reproduce previously established results in specific limiting cases. 
(i) In the absence of finite-range effects (i.e. $r_{\text{s}}=0$), we double check that our result aligns with Refs.~\cite{Song2022,Van2022} in terms of the gas parameter $\rho a_{\text{s}}^{3}$. 
Firstly, in the limit where $\rho a_{\text{s}}^{3}\ll1$, Eq.~(\ref{the solution of $M$}) decouples from the gas parameter $\rho a_{\text{s}}^{3}$, contributing to the LHY term in the universal EOS. 
Secondly, under the additional constraint $\rho a_{\text{s}}^{3}\ll\sqrt{\rho a_{\text{s}}^{3}}$, Eq.~(\ref{the solution of $M$}) truncates to the first order of $\sqrt{\rho a_{\text{s}}^{3}}$, contributing to beyond-LHY terms in the universal EOS.
(ii) When the finite-range effects are considered (i.e. $r_{\text{s}}\neq0$), our results accurately recover those in Ref.~\cite{Tononi2018} when calculating the EOS of nonuniversal systems using the expression for $M$.
This validation underscores the accuracy and applicability of our approach.

Having acquired the knowledge of $M$ in Eq. (\ref{the solution of $M$}), we are now poised to compute EOS of the model system, leveraging the CJT effective potential as presented in Eq. (\ref{the final Veff}). 
Specifically, the EOS explored in this paper pertains to two key quantities: the quantum depletion, denoted as $\rho_{\text{ex}}$, and chemical potential, represented by $\mu$.  

Using Equation~(\ref{the thermodynamic relation of rho}), the expression for quantum depletion is given by $\rho_{\text{ex}}=\frac{\sqrt{2}m^{*3/2}M^{3}}{12\pi^{2}\hbar^{3}}\left(2\sqrt{\frac{m^{*}}{m}}-\sqrt{\frac{m}{m^{*}}}\right)$.
By substituting the solution for $M$ from Eq.~(\ref{the solution of $M$}) into $\rho_{\text{ex}}$, we derive the analytical expression for the quantum depletion expanded in terms of the gas parameter $\rho a_{\text{s}}^{3}$
\begin{eqnarray}
\label{quantum depletion}
\rho_{\text{ex}} =&~&\!\!\!\!\!\!\frac{8\rho}{3\sqrt{\pi}}\left(\rho a_{\text{s}}^{3}\right)^{1/2}-\frac{128\rho}{3\pi}\left(\rho a_{\text{s}}^{3}\right)+\frac{2048\rho}{9\pi^{3/2}}\left(\rho a_{\text{s}}^{3}\right)^{3/2}\nonumber\\
 &-&64\sqrt{\pi}\rho\frac{r_{\text{s}}}{a_{\text{s}}}\left(\rho a_{\text{s}}^{3}\right)^{3/2}+\frac{5120}{3}\rho\frac{r_{\text{s}}}{a_{\text{s}}}\left(\rho a_{\text{s}}^{3}\right)^{2}.
\end{eqnarray}

Next, integrating the gap equation from Eq. (\ref{mu gap equation with P11 and P22}) 
with the thermodynamic relation for the density in Eq.~(\ref{the thermodynamic relation of rho}),
we systematically derive the analytical expression for the chemical potential.
The expression for the chemical potential takes the form $\mu=g_{0}\rho+g_{0}\frac{\sqrt{2}m^{*3/2}M^{3}}{3\pi^{2}\hbar^{3}}\sqrt{\frac{m^{*}}{m}}$.
By plugging the expression for $M$ from Eq.~(\ref{the solution of $M$}) into the formula of chemical potential, we are able to express 
$\mu$ as an expansion in terms of the gas parameter $\rho a_{\text{s}}^{3}$
\begin{eqnarray}
\label{mu}
\mu\!=\!g_{0}\rho\Big[1&+&\frac{32}{3\sqrt{\pi}}\left(\rho a_{\text{s}}^{3}\right)^{1/2}\!-\!\frac{512}{3\pi}\left(\rho a_{\text{s}}^{3}\right)\!+\!\frac{8192}{9\pi^{3/2}}\left(\rho a_{\text{s}}^{3}\right)^{3/2}\nonumber \\
 & -& \frac{512}{3}\sqrt{\pi}\frac{r_{\text{s}}}{a_{\text{s}}}\left(\rho a_{\text{s}}^{3}\right)^{3/2}+\frac{16384}{3} \frac{r_{\text{s}}}{a_{\text{s}}}\left(\rho a_{\text{s}}^{3}\right)^{2}\Big].
\end{eqnarray}

Equations (\ref{quantum depletion}) and (\ref{mu}) embody the key analytical expressions for nonuniversal EOS of a weakly-interacting Bose gas, with finite-range effects taken into consideration.
Our results of Eqs. (\ref{quantum depletion}) and (\ref{mu}), demonstrate their versatility by successfully recovering previously established results under specific limiting conditions.
(i) In the absence of the finite-range effects (i.e. $r_{\text{s}}=0$), 
our results align perfectly with the Refs.~\cite{Yang1957,Lee1957,Song2022,Van2022}, as verified through the order of the gas parameter $\rho a_{\text{s}}^{3}$. 
First, at the leading order $\left(\rho a_{\text{s}}^{3}\right)^{0}$, all particles condense at the mean field $\phi_{0}$,
resulting in zero excess density ($\rho_{\text{ex}}=0$) and chemical potential $\mu=g_{0}\rho$.
Second, proceeding to the next order $\left(\rho a_{\text{s}}^{3}\right)^{1/2}$, Eq.~(\ref{mu}) recovers the universal LHY term \cite{Yang1957,Lee1957}.
Third, truncating at the order of $\left(\rho a_{\text{s}}^{3}\right)^{3/2}$ for small gas parameter values, 
our results coincide with those reported in Ref.~\cite{Van2022} as they should be, 
emphasizing the consistency and robustness of our approach.
(ii) Then, when the finite-range effects are taken into account (i.e. $r_{\text{s}}\neq0$), 
our results maintain their consistency with the Refs.~\cite{Cappellaro2017,Tononi2018,Salasnich2017}, expanded in terms of the gas parameter $\rho a_{\text{s}}^{3}$. 
Notably, truncating within the order of $\left(\rho a_{\text{s}}^{3}\right)^{3/2}$, 
the nonuniversal EOS in Eqs.~(\ref{quantum depletion}) 
and (\ref{mu}) match precisely with those in Refs.~\cite{Cappellaro2017,Tononi2018,Salasnich2017}, obtained via the functional path-integral method.
Crucially, our approach offers a significant enhancement in precision. 
Not only do we approximate the universal EOS in Eqs.~(\ref{quantum depletion}) 
and (\ref{mu}) up to the order of $\left(\rho a_{\text{s}}^{3}\right)^{3/2}$, 
but also uncover the nonuniversal next-to-LHY EOS, given by $[16384r_{\text{s}}/3a_{\text{s}}]\left(\rho a_{\text{s}}^{3}\right)^{2}$, utilizing the CJT effective field theory.
This additional insight underscores the robustness and precision of our analytical framework.

In Equations~(\ref{quantum depletion}) and~(\ref{mu}), the EOS comprises two distinct components: the universal terms, described by the scattering length $a_{\text{s}}$, and the nonuniversal terms, described by both the scattering length $a_{\text{s}}$ and the finite-range length $r_{\text{s}}$ based on Lagrangian density~(\ref{Lagrangian density}). However, in Refs.~\cite{Wu1959,Braaten2001,Braaten1999}, the EOS has been calculated with both two-body correlations and three-body correlations
taken into account, which we have mentioned in the Introduction. In particular, the universal next-to-LHY term of energy density induced by three-body correlations is $\varepsilon=\frac{g_0 \rho^2}{2}\left[\frac{8\left(4\pi-3\sqrt{3}\right)}{3}\rho a_{\text{s}}^{3}\ln \left(\rho a_{\text{s}}^{3}\right)\right]$, as shown in Table~\ref{table1}.
The relative hamiltonian density can be written as $\mathcal{H}=\frac{\hbar^2}{2m}\nabla \mathbf{\Phi}^{*}\cdot\nabla\mathbf{\Phi}-\mu\left|\mathbf{\Phi}\right|^{2}
+\frac{U_0}{2}\left|\mathbf{\Phi}\right|^{4}+\frac{W}{6}\left|\mathbf{\Phi}\right|^{6}$.
In the tight-bingding limit, the relationship between dimensionless $\overline{W}$ and $\overline{U}_0$ is $\overline{W}=\left(3\pi\right)^{-3/2}\ln\left(C\eta^2\right)\left(\frac{V_0}{E_{\text{r}}}\right)^{3/4}
e^{-2\sqrt{\frac{V_0}{E_{\text{r}}}}}a_{\text{s}}^{2}
k^{2}\overline{U}^2_0$ \cite{Chen2008,Zhou2010}. Here, $\eta=\sqrt{\rho a_{\text{s}}^3}$, $C$ constant has been given in Ref.~\cite{Thorsten2002}, $V_0$ represents the tunable barrier height of a homogeneous periodic lattice potential, while $k$ denotes the wave vector and
$E_{\text{r}}=\hbar^2 k^2/2m$ is the recoil energy.
In contemporary experiments, $a_{\text{s}}^{2}k^{2}$ typically spans from $10^{-8}$ to $10^{-2}$~\cite{Pethick_Smith_2008}. Within this range, the influence of three-body interactions is markedly negligible in comparison to two-body interactions, implying that the effects stemming from three-body interactions are exceedingly difficult to discern experimentally.
Therefore, we neglect the three-body correlations when we calculate the EOS for our boson-boson interacting system.

  Meanwhile, the known next-to-LHY term of EOS induced by three-body correlations resides at the order of $\left(\rho a_{\text{s}}^{3}\right)^1$, aligning with the first term of universal next-to-LHY EOS though CJT effective field theory, as demonstrated in the third line in Table~\ref{table1}. It can be roughly interpreted that the CJT effective field theory essentially effects a higher-order perturbation of the two-body interacting system in comparison to a direct consideration of three-body interactions.
We note that the EOS in Eq.~(\ref{mu}) is at the order of $\left(\rho a_{\text{s}}^3\right)^2$,
which is higher than the results in Refs.~\cite{Braaten2001,Braaten1999}.
In this sense, we need to consider the three-body correlations. Meanwhile, there is no need to use CJT effective field theory when calculating the nonuniversal EOS induced by three-body effect, which is supposed to give the higher order than $\left(\rho a_{\text{s}}^{3}\right)^{2}$. It is enough to obtain a proper EOS through adding the terms induced by three-body correlations in Refs.~\cite{Braaten2001,Braaten1999} into our results in Eq. (\ref{mu}) by hand. We conclude our result together with the logarithmic term of the next-to-LHY correction to the universal EOS gives a reasonable description of the EOS of an ultracold Bose gas.

\begin{table*}
\renewcommand\arraystretch{1.7}
\caption{\label{table1}Expressions of the EOS expanded as the gas parameter $\rho a_{\text{s}}^{3}$ for weakly-interacting Bose gas. Here, $\mu$ signifies the chemical potential, $\rho_{\text{ex}}$ represents the quantum depletion and $\varepsilon$ denotes the energy density with $\rho$ being the density, $a_{\text{s}}$ being the $s$-wave scattering length and $r_{\text{s}}$ representing the finite-range length respectively. For three-body correlations, $l_{\text{V}}$ is the length scale set by the van der Waals potential, while the parameters $c_{\text{E}}$ and $b$ represent the nonuniversal and universal coefficients respectively. The nonuniversal beyond-LHY EOS of the Lagrangian density functional~(\ref{Lagrangian density}) is calculated with two-body correlations taken into account in this work. It gives a higher order term of $\rho_{\text{ex}}$ compared with the Ref.~\cite{Tononi2018}.}
\begin{ruledtabular}
\begin{tabular}{ccccc}
 Interatomic potential&Quantum fluctuation&
 EOS &Gas parameter$\left(\rho a_{\text{s}}^{3}\right)$&Ref.\\ \hline
 Two-body &  &\makecell[c]{$\mu=g_0\rho\big[1+\frac{32}{3\sqrt{\pi}}\left(\rho a_{\text{s}}^{3}\right)^{1/2}$\\
 $-\frac{512}{3\pi}\left(\rho a_{\text{s}}^{3}\right)+\!\frac{8192}{9\pi^{3/2}}\left(\rho a_{\text{s}}^{3}\right)^{3/2}$\\
 $-\frac{512}{3}\sqrt{\pi}\frac{r_{\text{s}}}{a_{\text{s}}}\left(\rho a_{\text{s}}^{3}\right)^{3/2}+\frac{16384}{3} \frac{r_{\text{s}}}{a_{\text{s}}}\left(\rho a_{\text{s}}^{3}\right)^{2}\big]$}&   &\textcolor{blue}{This work} \\ \hline
Two-body &Universal  &$\mu_{\text{MF}}=g_0\rho$&Mean-field term&Ref.~\cite{Dalfovo1999} \\
        &Universal     &$\mu_{\text{LHY}}=g_{0}\rho\big[\frac{32}{3\sqrt{\pi}}\left(\rho a_{\text{s}}^{3}\right)^{1/2}\big]$ &LHY term  &Ref.~\cite{Lee1957}\\
        &Universal     &$\mu_{\text{BLHY}}=g_{0}\rho\big[-\frac{512}{3\pi}\left(\rho a_{\text{s}}^{3}\right)+\!\frac{8192}{9\pi^{3/2}}\left(\rho a_{\text{s}}^{3}\right)^{3/2}\big]$&Beyond-LHY terms&Ref.~\cite{Van2022}\\
        Three-body &Universal&$\varepsilon_{\text{BLHY}}=\frac{g_0 \rho^2}{2}\big[\frac{8\left(4\pi-3\sqrt{3}\right)}{3}\rho a_{\text{s}}^{3}\ln \left(\rho a_{\text{s}}^{3}\right)\big]$&Beyond-LHY term&Ref.~\cite{Wu1959}\\
                               &Nonuniversal&$\varepsilon_{\text{BLHY}}=\frac{g_0 \rho^2}{2}\big[\frac{8\left(4\pi-3\sqrt{3}\right)}{3}\rho a_{\text{s}}^{3}\ln \left(\rho a_{\text{s}}l_{\text{V}}^2\right)\big]$&Beyond-LHY term&Ref.~\cite{Braaten1999}\\
                               &Nonuniversal&
                               \makecell{$\varepsilon_{\text{BLHY}}=\frac{g_0 \rho^2}{2}\Big\{\big[c_{\text{E}}+\frac{\left(4\pi-3\sqrt{3}\right)}{6\pi}\ln \left(16\pi\rho a_{\text{s}}^{3}\right)+\frac{4}{9\pi^{2}}\big]$\\
                               $\left(16\pi\rho a_{\text{s}}^{3}\right)+\Big(\frac{16}{\pi}[c_{\text{E}}+\frac{\left(4\pi-3\sqrt{3}\right)}{6\pi}\ln \left(16\pi\rho a_{\text{s}}^{3}\right)]$\\
                               $-\frac{16}{15\pi}\frac{r_{\text{s}}}{a_{\text{s}}}+b\Big)\left(16\pi\rho a_{\text{s}}^{3}\right)^{3/2}\Big\}$
                               }                 &Beyond-LHY terms&Ref.~\cite{Braaten2001}\\
Two-body         &Nonuniversal       &$\rho^{\text{LHY}}_{\text{ex}}=-64\sqrt{\pi}\rho\frac{r_{\text{s}}}{a_{\text{s}}}\left(\rho a_{\text{s}}^{3}\right)^{3/2}$&LHY term&Ref.~\cite{Tononi2018}\\
        &Nonuniversal            &$\rho^{\text{NLHY}}_{\text{ex}}=\frac{5120}{3}\rho\frac{r_{\text{s}}}{a_{\text{s}}}\left(\rho a_{\text{s}}^{3}\right)^{2}$&Next-to-LHY term&\textcolor{blue}{This work}\\
\end{tabular}
\end{ruledtabular}
\end{table*}

Next, based on the chemical potential $\mu$, the inverse compressibility $\kappa^{-1}=\rho\frac{\partial\mu}{\partial\rho}$ can be readily derived
\begin{eqnarray}
\label{kappa}
\kappa^{-1}=&~&\!\!\!\!\!\!g_{0}\rho\Big[1+\frac{3}{2}\alpha\rho^{1/2}-3\alpha^{2}\rho+\frac{15}{8}\alpha^{3}\rho^{3/2}\nonumber \\
&-&\frac{45\pi^{2}}{128}\frac{r_{\text{s}}}{a_{\text{s}}}\alpha^{3}\rho^{3/2}+\frac{81\pi^{2}}{64}\frac{r_{\text{s}}}{a_{\text{s}}}\alpha^{4}\rho^{2}\Big],
\end{eqnarray}
with $\alpha=\left(32/3\sqrt{\pi}\right)a_{\text{s}}^{3/2}$.

To delve deeper into our analysis, utilizing the Eq.~(\ref{kappa}), we explore the nonuniversal quantum effect present in the speed of sound, denoted as $c_{\text{s}}=\sqrt{\left(\kappa m\right)^{-1}}$, 
\begin{eqnarray}
\label{cs}
c_{\text{s}}=\!\!\sqrt{\frac{g_{0}\rho}{m}}\Bigg\{&1&+\frac{8\sqrt{\rho a_{\text{s}}^{3}}}{\sqrt{\pi}}\Big[1-\frac{64\sqrt{\rho a_{\text{s}}^{3}}}{3\sqrt{\pi}}+\frac{1280}{9\pi}\rho a_{\text{s}}^{3}\nonumber\\
&-&\frac{80\pi}{3}\frac{r_{\text{s}}}{a_{\text{s}}}\rho a_{\text{s}}^{3}+1024\sqrt{\pi}\frac{r_{\text{s}}}{a_{\text{s}}}\sqrt{\rho a_{\text{s}}^{3}}^{3}\Big]\Bigg\}.
\end{eqnarray}

In Equation~(\ref{cs}), the coefficient associated with the term $\sqrt{\rho a_{\text{s}}^{3}}$ is $8/\sqrt{\pi}$, 
differing from LHY's prediction of $16/\sqrt{\pi}$. 
The deviation is readily traceable to the corresponding corrective terms present in the definition of the inverse compressibility $\kappa^{-1}$, given by Eq.~(\ref{kappa}).
As a consequence of the nonuniversal effect, our derived expression for the sound speed $c_{\text{s}}$ incorporates higher-order terms in the gas parameter $\rho a_{\text{s}}^{3}$,
distinguishing it from the findings presented in Ref.~\cite{Song2022}. 
This feature underscores the paramount importance and significance of our approach.
\section{Frequency shifts induced by NONUNIVERSAL beyond-LHY EOS}\label{CETN}
In Sec.~\ref{NEST}, we have rigorously derived the nonuniversal next-to-LHY term in EOS (see Eqs. (\ref{quantum depletion}) and (\ref{mu}))
to the order of $\left(\rho a_{\text{s}}^{3}\right)^2$. 
The purpose of Sec.~\ref{CETN} is to propose an experimental protocol to observe the nonuniversal beyond-LHY corrections to the EOS by calculating the frequency shifts in the breathing modes.
In contemporary experiments, the gas parameter $\rho a_{\text{s}}^{3}$ typically remains below $10^{-4}$,
rendering even the initial LHY correction to the mean-field energy, scaling as $\sqrt{\rho a_{\text{s}}^3}$,
negligible at best, contributing less than $1\%$ of the total energy.
Consequently, the higher-order corrections beyond LHY in Eq.~(\ref{mu}) 
are anticipated to be too subtle to discern in density profiles or release energies. 
While enhancing quantum fluctuations by tuning $a_{\text{s}}$ via Feshbach resonance~\cite{Cornish_2000,Pethick_2002,TB_2004,Chin2010,Wu2012,HaibinWU2012} can amplify their impact, 
we propose an alternative route: studying the frequency shifts in collective excitations~\cite{AGE_2005,Liang_2010,Hu2011,Yun_2011} induced by quantum fluctuations~\cite{Hu2009}.
To this end, we augment our model system with a 3D harmonic trap, defined by $V_{\text{ext}}=\frac{1}{2}m\omega^2_0r^2$,
and embark on calculating the resulting shifts in the breathing mode frequency, attributed to beyond-LHY terms in Eq.~(\ref{mu}). 
This endeavor is motivated by the potential to experimentally verify and quantify the nonuniversal contributions to the EOS.

Adhering closely to the standard methodology outlined in Ref.~\cite{Pitaevskii1998} for calculating frequency shifts, 
we formulate the hydrodynamic equation as follows:
\begin{equation}\label{hydrodynamic equation}
 m\frac{\partial^{2}\delta \rho\left(\mathbf{r},t\right)}{\partial t^{2}}-\nabla\cdot\left[\rho\left(\mathbf{r}\right)\nabla\left(\frac{\partial\mu_{l}}{\partial\rho}\delta \rho\left(\mathbf{r},t\right)\right)\right]=0.
\end{equation}
Equation (\ref{hydrodynamic equation}) incorporates the density fluctuation, denoted as $\delta \rho\left(\mathbf{r},t\right)$, 
which arises around the targeted ground state density
$\rho\left(\mathbf{r}\right)$ and the local chemical potential $\mu_{l}$ 
as specified in Eq.~(\ref{mu}).

As an initial step, utilizing Eq.~(\ref{mu}), we can iteratively derive the expansion of the ground state density, 
expressed in terms of $\alpha=\left(32/3\sqrt{\pi}\right)a_{\text{s}}^{3/2}$,
\begin{eqnarray}
\label{rho expansion of rhoTF}
\rho\left(\mathbf{r}\right)= &~&\!\!\!\!\!\! \rho_{\text{TF}}-\alpha\rho_{\text{TF}}^{3/2}+\frac{3}{2}\alpha^{2}\rho_{\text{TF}}^{2}-\frac{3}{4}\alpha^{3}\rho_{\text{TF}}^{5/2}\nonumber \\
 &+&\frac{9}{64}\pi^{2}\frac{r_{\text{s}}}{a_{\text{s}}}\alpha^{3}\rho_{\text{TF}}^{5/2}-\frac{27}{64}\pi^{2}\frac{r_{\text{s}}}{a_{\text{s}}}\alpha^{4}\rho_{\text{TF}}^{3}.
\end{eqnarray}
In Equation~(\ref{rho expansion of rhoTF}), $\rho_{\text{TF}}=\left(g_{0}\rho-V_{\text{ext}}\right)/g_{0}$ \cite{Stringari1996,Pitaevskii1998}, 
is the so-called Thomas-Fermi result for the ground state density.
Examining Eq.~(\ref{rho expansion of rhoTF}), we observe that its first line comprises three distinct terms: 
the mean-field term being $\rho_{\text{TF}}$, the universal LHY term represented by $-\alpha\rho_{\text{TF}}^{3/2}$, 
and the universal beyond-LHY term $3/2\alpha^{2}\rho_{\text{TF}}^{2}-3/4\alpha^{3}\rho_{\text{TF}}^{5/2}$.
Notably, the second line of Eq.~(\ref{rho expansion of rhoTF}) vanishes solely when nonuniversal effects are disregarded, 
emphasizing their significance. Consequently, $\frac{9}{64}\pi^{2}\frac{r_{\text{s}}}{a_{\text{s}}}\alpha^{3}\rho_{\text{TF}}^{5/2}$
is designated as the nonuniversal LHY term, while
$-\frac{27}{64}\pi^{2}\frac{r_{\text{s}}}{a_{\text{s}}}\alpha^{4}\rho_{\text{TF}}^{3}$
corresponds to the nonuniversal next-to-LHY term. 

Subsequently, we derive the expansion of $\left[\rho \partial \mu_{l}/\partial \rho\right]$ as a series in terms of $\rho_{\text{eff}}$
by substituting Eq.~(\ref{rho expansion of rhoTF}) into Eq.~(\ref{kappa}). Following this substitution, 
we insert both Eqs.~(\ref{kappa}) and~(\ref{rho expansion of rhoTF}) into the Eq.~(\ref{hydrodynamic equation}).  
Through meticulous algebraic manipulations, we ultimately arrive at 
\begin{widetext}
 \begin{eqnarray}
 \label{hydrodynamic equation & the expansion of rhoTF}
\!\!\!\!\!\!\!\!m\omega^{2}\delta\rho  +\nabla\cdot\left[g_{0}\rho_{\text{TF}}\nabla \delta\rho\right]=&-&\frac{1}{2}\alpha\nabla^{2}\left(g_{0}\rho_{\text{TF}}^{3/2}\delta\rho\right)+\frac{15}{4}\alpha^{2}\nabla^{2}\left(g_{0}\rho_{\text{TF}}^{2}\delta\rho\right)-
3\alpha^{2}\nabla\cdot\left(g_{0}\rho_{\text{TF}}\delta\rho\nabla\rho_{\text{TF}}\right)\nonumber \\
&+&\left(-\frac{45}{4}+\frac{63\pi^{2}}{128}\frac{r_{\text{s}}}{a_{\text{s}}}\right)\alpha^{3}\nabla^{2}\left(g_{0}\rho_{\text{TF}}^{5/2}\delta\rho\right)
+\frac{57}{4}\alpha^{3}\nabla\cdot\left(g_{0}\rho_{\text{TF}}^{3/2}\delta\rho\nabla\rho_{\text{TF}}\right)\nonumber \\
&+&\left(\frac{165}{8}-\frac{261\pi^{2}}{128}\frac{r_{\text{s}}}{a_{\text{s}}}\right)\alpha^{4}\nabla^{2}\left(g_{0}\rho_{\text{TF}}^{3}\delta\rho\right)
+\left(-\frac{495}{16}+\frac{27\pi^{2}}{16}\frac{r_{\text{s}}}{a_{\text{s}}}\right)\alpha^{4}\nabla\cdot\left(g_{0}\rho_{\text{TF}}^{2}\delta\rho\nabla\rho_{\text{TF}}\right).
\end{eqnarray}
\end{widetext}
Equation (\ref{hydrodynamic equation & the expansion of rhoTF}) simplifies into the basic hydrodynamic equation
$m\omega^{2}\delta\rho+\nabla\cdot\left[g_{0}\rho_{\text{TF}}\nabla \delta\rho\right]=0$ in case of $\alpha=0$.
For this simplified scenario, the calculated frequency $\omega$ exhibits analytical solutions of the form
$\omega\left(n_{r},l\right)=\omega_{0}\left(2n_{r}^{2}+2n_{r}l+3n_{r}+l\right)^{1/2}$, 
with $n_{r}$ representing the number of radial nodes and $l$ denoting the angular momentum associated with the excitation.
This analytical expression provides a direct link between the frequency of the excitation and its quantum numbers.

Finally, Equation~(\ref{hydrodynamic equation & the expansion of rhoTF}) can be routinely tackled 
by considering its right-hand side as a minor perturbation. 
By adopting this approach, we can derive the analytical expression for the frequency shifts, 
\begin{widetext}
\begin{eqnarray}
\label{delta omega/omega}
\frac{\delta\omega}{\omega}= &-&\frac{\alpha g_{0}}{4m\omega^{2}}\frac{\int d^{3}{\mathbf r}\nabla^{2}\left(\delta\rho^{*}\right)\rho_{\text{TF}}^{3/2}\delta\rho}{\int d^{3}{\mathbf r}\delta\rho^{*}\delta\rho}+\frac{15\alpha^{2}g_{0}}{8m\omega^{2}}\frac{\int d^{3}{\mathbf r}\nabla^{2}\left(\delta\rho^{*}\right)\rho_{\text{TF}}^{2}\delta\rho}{\int d^{3}{\mathbf r}\delta\rho^{*}\delta\rho}
+\frac{3\alpha^{2}g_{0}}{2m\omega^{2}}\frac{\int d^{3}{\mathbf r}\nabla\left(\delta\rho^{*}\right)\cdot\nabla\left(\rho_{\text{TF}}\right)\rho_{\text{TF}}\delta\rho}{\int d^{3}{\mathbf r}\delta\rho^{*}\delta\rho}\nonumber \\
&+&\frac{\left(-\frac{45}{4}+\frac{63\pi^{2}}{128}\frac{r_{\text{s}}}{a_{\text{s}}}\right)\alpha^{3}g_{0}}{2m\omega^{2}}\frac{\int d^{3}{\mathbf r}\nabla^{2}\left(\delta\rho^{*}\right)\rho_{\text{TF}}^{5/2}\delta\rho}{\int d^{3}{\mathbf r}\delta\rho^{*}\delta\rho}
-\frac{\frac{57}{4}\alpha^{3}g_{0}}{2m\omega^{2}}\frac{\int d^{3}{\mathbf r}\nabla\left(\delta\rho^{*}\right)\cdot\nabla\left(\rho_{\text{TF}}\right)\rho_{\text{TF}}^{3/2}\delta\rho}{\int d^{3}{\mathbf r}\delta\rho^{*}\delta\rho}\nonumber\\
&+&\frac{\left(\frac{165}{8}-\frac{261\pi^{2}}{128}\frac{r_{\text{s}}}{a_{\text{s}}}\right)\alpha^{4}g_{0}}{2m\omega^{2}}\frac{\int d^{3}{\mathbf r}\nabla^{2}\left(\delta\rho^{*}\right)\rho_{\text{TF}}^{3}\delta\rho}{\int d^{3}{\mathbf r}\delta\rho^{*}\delta\rho}
-\frac{\left(-\frac{495}{16}+\frac{27\pi^{2}}{16}\frac{r_{\text{s}}}{a_{\text{s}}}\right)\alpha^{4}g_{0}}{2m\omega^{2}}\frac{\int d^{3}{\mathbf r}\nabla\left(\delta\rho^{*}\right)\cdot\nabla\left(\rho_{\text{TF}}\right)\rho_{\text{TF}}^{2}\delta\rho}{\int d^{3}{\mathbf r}\delta\rho^{*}\delta\rho}.
\end{eqnarray}
\end{widetext}
The integrals in Eq.~(\ref{delta omega/omega}) are confined to the domain where the Thomas-Fermi density remains positive. 
To delve into effects that transcend the LHY theory, it becomes imperative to concentrate on compressional modes, 
which are highly responsive to alterations in the EOS. 
Among these, the breathing mode in a spherical trap, characterized by $\left(n_r=1,l=0\right)$,
stands as the fundamental excitation. This mode is distinguished by its zeroth-order dispersion, yielding a frequency of 
$\omega=\sqrt{5}\omega_0$ and exhibits density oscillations that adhere to the pattern 
$\delta\rho\sim\left(r^2-3/5R^2\right)$. 
In this specific scenario, Eq.~(\ref{delta omega/omega}) furnishes:
\begin{eqnarray}
\label{the result of delta omega/omega} 
\frac{\delta\omega}{\omega}=&~&\!\!\!\!\!\! \frac{63\sqrt{\pi}}{128}\left[\rho\left(0\right)a_{\text{s}}^{3}\right]^{1/2}-\frac{16384}{135\pi}\left[\rho \left(0\right)a_{\text{s}}^{3}\right]\nonumber \\
 &+&\frac{3682-\frac{2205}{16}\pi^{2}\frac{r_{\text{s}}}{a_{\text{s}}}}{3\sqrt{\pi}}\left[\rho\left(0\right)a_{\text{s}}^{3}\right]^{3/2}\nonumber \\
 &+&\left(\frac{-71200}{\pi^{2}}+6465\frac{r_{\text{s}}}{a_{\text{s}}}\right)\left[\rho\left(0\right)a_{\text{s}}^{3}\right]^{2},
\end{eqnarray}
revealing the fractional shift in the breathing mode frequency, where $\rho\left(0\right)$ represents the density evaluated
at the center of the trap.

Equation~(\ref{the result of delta omega/omega}) constitutes another pivotal finding, 
encapsulating corrections to the breathing mode frequency 
that scale with the gas parameter $\rho\left(0\right)a_{\text{s}}^{3}$.
The corrections in Eq.~(\ref{the result of delta omega/omega}) stem from diverse origins, encompassing both LHY and beyond-LHY contributions to the EOS. 
The first term on the right-hand side of Eq.~(\ref{the result of delta omega/omega}) represents the ubiquitous LHY-mediated fractional shift of the breathing mode
frequency, which scales as $\left[\rho\left(0\right)a_{\text{s}}^{3}\right]^{1/2}$.
All the subsequent terms in Eq.~(\ref{the result of delta omega/omega})
are of higher order than $\left[\rho\left(0\right)a_{\text{s}}^{3}\right]^{1/2}$, offering insights into more intricate effects. 
Notably, among these beyond-LHY terms, the one scaling with $\left[\rho \left(0\right)a_{\text{s}}^{3}\right]$ is 
a so-called next-to-LHY universal fractional shift. 
This term, specifically $\frac{-16384}{135\pi}\left[\rho \left(0\right)a_{\text{s}}^{3}\right]$,
underscores the persistence of universal behavior beyond the leading LHY correction. Particularly, terms in Eq.~(\ref{the result of delta omega/omega}) involving 
$r_\text{s}$ represent nonuniversal contributions, arising from the finite-range effective potential, 
further illuminating the intricacies of the system's response.
Thus, the complexity deepens with the term $\frac{3682-\frac{2205}{16}\pi^{2}\frac{r_{\text{s}}}{a_{\text{s}}}}{3\sqrt{\pi}}\left[\rho\left(0\right)a_{\text{s}}^{3}\right]^{3/2}$
in Eq.~(\ref{the result of delta omega/omega}),
which encapsulates both the next-next-to-LHY universal fractional shift  
$\frac{3682}{3\sqrt{\pi}}\left[\rho\left(0\right)a_{\text{s}}^{3}\right]^{3/2}$ 
and a LHY nonuniversal contribution 
$\frac{-\frac{2205}{16}\pi^{2}\frac{r_{\text{s}}}{a_{\text{s}}}}{3\sqrt{\pi}}\left[\rho\left(0\right)a_{\text{s}}^{3}\right]^{3/2}$.
Similarly, the last term $\left(\frac{-71200}{\pi^{2}}+6465\frac{r_{\text{s}}}{a_{\text{s}}}\right)\left[\rho\left(0\right)a_{\text{s}}^{3}\right]^{2}$
in Eq.~(\ref{the result of delta omega/omega}) 
comprises two distinct components: the next-next-next-to-LHY universal fractional shift 
$\frac{-71200}{\pi^{2}}\left[\rho\left(0\right)a_{\text{s}}^{3}\right]^{2}$ 
and the next-to-LHY nonuniversal fractional shift 
$6465\frac{r_{\text{s}}}{a_{\text{s}}}\left[\rho\left(0\right)a_{\text{s}}^{3}\right]^{2}$. 
We remark that to our best knowledge, 
the analytical expressions of beyond-LHY-induced fractional shift of the breathing mode frequency in 
Eq.~(\ref{the result of delta omega/omega}) are obtained for the first time.

Before delving into the intricacies concerning the fractional shift in the breathing mode frequency 
as described in Eq.~(\ref{the result of delta omega/omega}),
it is crucial to assess the rationality of the dimensionless finite-range coupling constant $r_{\text{s}}/a_{\text{s}}$ 
and the gas parameter 
$\rho\left(0\right)a_{\text{s}}^{3}$. 
This preliminary examination is fundamental as it sheds light on the experimental feasibility of our proposed model. 
To exemplify, let us take the case of $^{6}\text{Li}$, as mentioned in Ref.~\cite{Bartenstein2005}. 
In this context, the typical density $\rho\left(0\right)$ is approximately $4\times 10^{12}\text{cm}^{-3}$. 
while the scattering length $a_{\text{s}}$ is estimated to be approximately $1.13\times 10^{-7} \text{m}$.
Consequently, the magnitude of $\rho\left(0\right)a_{\text{s}}^{3}$ falls within the order of $10^{-3}$. 
Furthermore, according to Ref.~\cite{Haibin2012}, the effective distance $r_{\text{s}}$ is estimated to lie within the interval of 
$0\sim\pm3.71\times 10^{-6}\text{m}$. 
By substituting these parameters into the expressions for
$r_{\text{s}}/a_{\text{s}}$ and $\rho\left(0\right)a_{\text{s}}^{3}$, 
we can reasonably infer that $r_{\text{s}}/a_{\text{s}}$ falls within the range of $0\sim\pm1$.
This assessment underscores the physical relevance and potential experimental applicability of our model's key parameters.
\begin{figure}[H]
  \centering
  \includegraphics[scale=0.5]{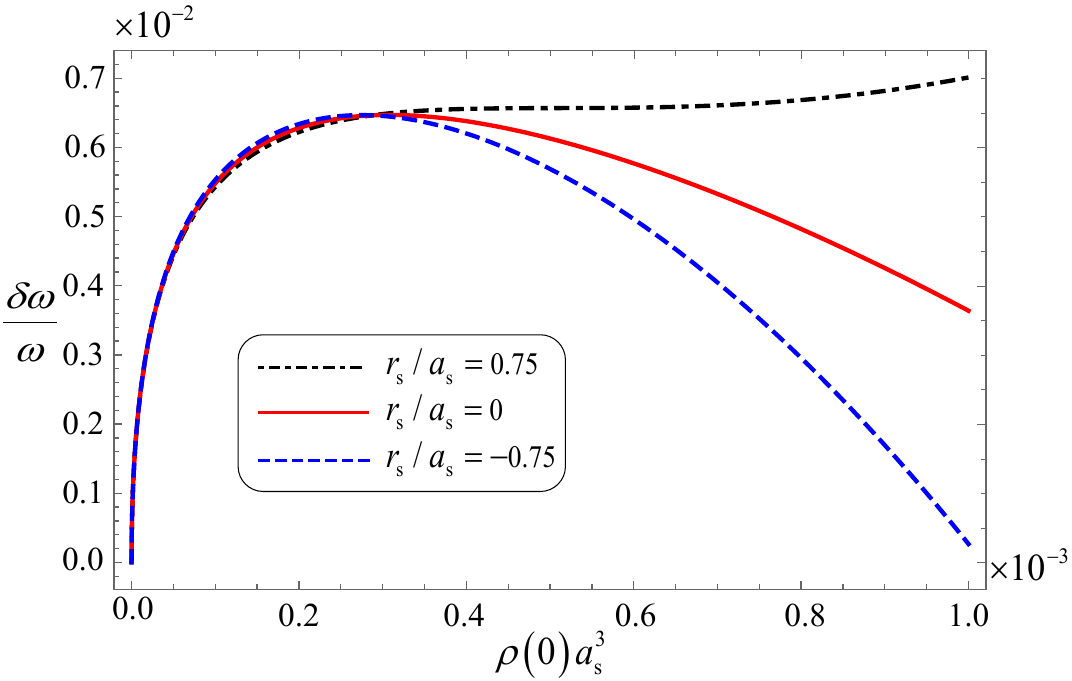}
  \caption{Frequency shifts $\delta\omega/\omega$ as a function of the gas parameter $\rho a_{\text{s}}^3$ with different values of $r_{\text{s}}/a_{\text{s}}$. All the curves are plotted by Eq.~(\ref{the result of delta omega/omega}), showing the frequency shifts in breathing modes arising from the nonuniversal beyond-LHY EOS of a 3D Bose gas.}\label{a}
\end{figure}
We are now poised to discuss the influence of the nonuniversal EOS in Eq.~(\ref{mu}), stemming from the finite-range interaction parameter 
$r_{\text{s}}$, on the fractional shift of the breathing mode frequency in Eq.~(\ref{the result of delta omega/omega}). To visualize this effect, we have plotted frequency shifts of the breathing mode $\delta\omega/\omega$ 
in Eq.~(\ref{the result of delta omega/omega})  
into Fig.~\ref{a}, showing how the dimensionless finite-range interaction parameter of
$r_{\text{s}}/a_{\text{s}}$ can affect the frequency shifts of $\delta\omega/\omega$. Note that the red solid curve in Fig.~\ref{a} corresponds to case of vanishing the finite-range interaction by taking the dimensionless parameter of $r_{\text{s}}/a_s=0$; meanwhile
the black dot-dashed and blue dashed curves in Fig.~\ref{a} demonstrate how the attractive ($r_{\text{s}}/a_{\text{s}}=0.75$) and repulsive ($r_{\text{s}}/a_{\text{s}}=-0.75$) finite-range interaction can affect frequency shifts of $\delta\omega/\omega$. As shown in Fig.~\ref{a}, the finite-range interaction has relative small effects on frequency shifts of $\delta\omega/\omega$ compared with the case of vanishing the finite-range interaction when the gas parameter of $\rho a^3_{\text{s}}$ is small. In contrast, with the increase of the $\rho a^3_{\text{s}}$, the effect of finite-range interaction on frequency shifts of $\delta\omega/\omega$ becomes to be significant. For example, at the typical experimental parameter onset of $\rho a^3_\text{s}\sim 10^{-3}$,  the derivation of the fractional shift of $\delta\omega/\omega$ in case of  $r_{\text{s}}/a_{\text{s}}=0.75$ from the case of $r_{\text{s}}/a_{\text{s}}=-0.75$ is calculated as more than $0.5\%$, showing the finite-range effect well in reach in experiments. We point out that a precision of $<0.3\%$ in measuring collective frequencies has already been established~\cite{Mordini2020,Skov2021,Lavoine2021,Cominotti2023,Science2022}, offering opportunities to probe the nonuniversial beyond-LHY corrections to EOS experimentally.

\section{CONCLUSION and outlook}\label{DAC}
In summary, we have theoretically investigated the nonuniversal  EOS for a weakly-interacting Bose gas with the finite-range interatomic interaction. 
With the framework of CJT effective field theory under the two-loop approximation,  
we obtain analytical expressions for quantum depletion and chemical potential of model system, representing the next-to-LHY correction to nonuniversal EOS induced by finite-range effects.
These analytical results represent significant generalizations of the nonuniversal LHY EOS studied 
in Refs. \cite{Salasnich2017, Cappellaro2017, Tononi2018} to the beyond-LHY regimes,
offering fresh insights into understanding the quantum behavior induced by the quantum fluctuations in many-body bosonic systems.
We further calculate the frequency shifts in the breathing mode induced by the nonuniversal beyond-LHY EOS. Therefore, the beyond-LHY effects studied in this work should 
be observable within the current experiment capability. 

We finally remark that the CJT theory developed in this work can be readily applied to other ultracold quantum systems, including the novel quantum droplet phases of interacting Bose mixtures or ultracold quantum Fermi systems. Our results lay the groundwork for further investigation of the nonuniversal beyond-LHY EOS.

We thank Xiaoran Ye, Tao Yu and Ying Hu for stimulating discussions. This work was supported by the National Natural Science Foundation of China (Nos. 12074344), the Zhejiang Provincial Natural Science Foundation (Grant Nos. LZ21A040001) and the key projects of the Natural Science Foundation of China (Grant No. 11835011).

\appendix
\section{CJT EFFECTIVE FIELD THEORY}\label{CEAA}
For the purpose of maintaining self-consistency within this work, we provide a concise overview 
of the key steps of the Cornwall-Jackiw-Tomboulis (CJT) effective field theory, 
drawing from seminal works (see e.g. Refs.~\cite{Cornwall1974,Amelino1993} and the references therein). 
Specifically, we delve into the detailed derivation of the effective potential $V_{\text{eff}}$ as presented in 
Eq.~(\ref{Veff_phi_0,P_11,P_22}), elucidating each step to ensure a comprehensive understanding.

To begin with,
let us contemplate the partition function, which incorporates both linear and bilinear external sources,
\begin{widetext}
\begin{align}
\label{Z,JK}
\mathcal{Z}\left[J,K\right] & =e^{-W\left[J,K\right]/\hbar}=\int\mathcal{D}\left[\mathbf{\Phi}\right]\exp\left\{-\frac{1}{\hbar}\left[\int d^{4}x\mathcal{L}\left[\mathbf{\Phi}\right]+J\left(x\right)\mathbf{\Phi}\left(x\right)+\frac{1}{2}\int d^{4}xd^{4}y\mathbf{\Phi}\left(x\right)K\left(x,y\right)\mathbf{\Phi}\left(y\right)\right]\right\}.
\end{align}
\end{widetext}
In Equation~(\ref{Z,JK}), $\int d^{4}x=\int_{0}^{\beta\hbar}d\tau\int d^{3}\mathbf{r}$.
So that we have
\begin{subequations}
\begin{eqnarray}
\frac{\delta W\left[J,K\right]}{\delta J\left(x\right)}&=&-\frac{\delta\ln\mathcal{Z}}{\delta J\left(x\right)}=\left\langle \mathbf{\Phi}\left(x\right)\right\rangle _{J,K}=\phi\left(x\right),\\
\frac{\delta W\left[J,K\right]}{\delta K\left(x,y\right)} & =&-\frac{\delta\ln\mathcal{Z}}{\delta K\left(x,y\right)}=\frac{1}{2}\left\langle \mathbf{\Phi}\left(x\right)\mathbf{\Phi}\left(y\right)\right\rangle _{J,K}\nonumber \\
 & =&\frac{1}{2}\left[\phi\left(x\right)\phi\left(y\right)+\hbar G\left(x,y\right)\right]\label{W/K}.
\end{eqnarray}
\end{subequations}

The effective action, denoted as $\Gamma\left[\phi,G\right]$,
is precisely defined through the application of the double Legendre transformation to the generating functional $W\left[J,K\right]$.
This transformation yields the expression:
\begin{eqnarray}
\label{gamma_phi,G from W}
\!\!\!\!&&\Gamma\left[\phi,G\right]=\!W\left[J,K\right]-\int d^{4}u\phi\left(u\right)J\left(u\right)\nonumber \\
 \!\!\!\!\!\!&~&-\frac{1}{2}\!\int\!\! d^{4}vd^{4}wK\left(v,w\right)\left[\phi\left(v\right)\phi\left(w\right)+\hbar G\left(v,w\right)\right],
\end{eqnarray}
which is subject to the conditions encapsulated in the following set of equations:
\begin{subequations}
\begin{eqnarray}
\frac{\delta\Gamma\left[\phi,G\right]}{\delta\phi\left(x\right)} & =&-J\left(x\right)-\int d^{4}wK\left(x,w\right)\phi\left(w\right),\label{delta gamma/delta phi}\\
\frac{\delta\Gamma\left[\phi,G\right]}{\delta G\left(x,y\right)} & =&-\frac{1}{2}\hbar K\left(x,y\right)\label{delta gamma/delta G}.
\end{eqnarray}
\end{subequations}
Equations~(\ref{delta gamma/delta phi}) and (\ref{delta gamma/delta G}) underscore the intricate relationship between the effective action and its functional derivatives 
with respect to the classical field $\phi$ and the two-point function $G$, respectively.
After the saddle-point approximation, we calculate $\mathcal{Z}$ as
\!
\begin{align}
\label{saddle point approximation Z}
\mathcal{Z} & \simeq e^{-S\left[\phi,J,K\right]/\hbar}\int\mathcal{D}[\widetilde{\mathbf{\Phi}}]e^{-\frac{1}{2\hbar}\ensuremath{\int}d^4xd^4y\tilde{\mathbf{\Phi}}\left(x\right)\left[D_{0}^{-1}+K\right]\tilde{\mathbf{\Phi}}\left(y\right)}\nonumber \\
 & =e^{-S\left[\phi,J,K\right]/\hbar}\left[\det\left[\left(D_{0}^{-1}+K\right)/\hbar\right]\right]^{-1/2},
\end{align}
with
\begin{equation}
\label{S,phi,J,K}
S\left[\phi,J,K\right] =\int d^{4}x\phi D_{0}^{-1}\phi+\int d^{4}xd^{4}y\phi \frac{K}{2}\phi+\int d^{4}xJ\phi.
\end{equation}
Plugging Eqs.~(\ref{S,phi,J,K}) and~(\ref{saddle point approximation Z}) into Eq.~(\ref{gamma_phi,G from W}), we can obtain
\begin{widetext}
\begin{align}
\label{gamma_phi,G}
\Gamma\left[\phi,G\right] & =\frac{1}{2}\int d^{4}\phi D_{0}^{-1}\phi+\frac{\hbar}{2}\text{Tr}\ln\left[\left(D_{0}^{-1}+K\right)\right]-\frac{\hbar}{2}\int d^{4}vd^{4}wK\left(v,w\right)G\left(v,w\right).
\end{align}
\end{widetext}
By taking the logarithm of Eq.~(\ref{saddle point approximation Z}) and 
subsequently computing its first-order derivative with respect to $K\left(x,y\right)$,
utilizing Eq.~(\ref{S,phi,J,K}) as an aid, we can establish a direct comparison with Eq.~(\ref{W/K}). 
This comparison leads to the conclusion that $G^{-1}=D_{0}^{-1}+K$. Consequently, $\Gamma\left[\phi,G\right]$ 
can be expressed as
\begin{eqnarray}
\label{gamma_phi,G+Phi}
\Gamma\left[\phi,G\right] &=&\frac{1}{2}\int d^{4}x\phi D_{0}^{-1}\phi\nonumber \\
 & +&\frac{\hbar}{2}\text{Tr}\left[\ln G^{-1}+D_{0}^{-1}G-\mathbf{1}\right]\nonumber \\
 & +&\mathcal{I}\left[\phi,G\right],
\end{eqnarray}
where $\mathcal{I}\left[\phi,G\right]$ is the Luttinger-Ward functional.
However, our primary focus lies solely on translation-invariant solutions. To this end, we simplify our analysis by setting 
$\phi\left(x\right)$ to a constant value, denoted as $\phi_0$, and considering $G\left(x,y\right)$ 
as a function exclusively dependent on the difference $x-y$. 
This specific choice allows us to define a generalized form of the effective potential as
\begin{eqnarray}
\label{V_CJT}
V_{\text{eff}} & =&\frac{1}{\beta\hbar\mathcal{V}}\Gamma\left[\phi_{0},G\right]\nonumber \\
 & =&\frac{1}{2}\phi_{0}D_{0}^{-1}\phi_{0}+\frac{1}{\beta\mathcal{V}}\left[ \frac{1}{2}\text{Tr}\left(\ln G^{-1}+D_{0}^{-1}G-\mathbf{1}\right)\right] \nonumber \\
 &~&+\Phi\left[\phi_{0},G\right].
\end{eqnarray}
with $\Phi\left[\phi_{0},G\right]$ being the average of time and space of the $\mathcal{I}\left[\phi_{0},G\right]$.
\section{DETAILED DERIVATION OF THE EFFECTIVE POTENTIAL $V_{\text{eff}}$ OF EQ.~(\ref{the final Veff})}\label{DDTE}
To ensure the self-consistency and comprehensiveness of this work, 
we provide a concise overview of the crucial steps involved in deriving the self-consistent $V_{\text{eff}}$
that fulfills the condition of gapless excitations (see e.g. Refs.~\cite{Ivanov2005,Van2022} and the references therein). 
In particular, we delve into the derivation of the effective potential $V_{\text{eff}}$ as presented in Eq.~(\ref{the final Veff}).
To effectively implement the CJT effective field theory \cite{Cornwall1974}, 
it is advantageous to initially transform the CJT effective potential into the momentum-frequency space
\begin{subequations}
\label{the fourier transformation of phi}
\begin{eqnarray}
\text{\ensuremath{\phi_{1}}\ensuremath{\left(\mathbf r,\tau\right)}} & =&\ensuremath{\sqrt{\frac{1}{\beta \hbar \mathcal{V}}}}\ensuremath{\sum_{{\mathbf k},\omega_{n}}e^{i{\mathbf k}\cdot {\mathbf r}-i\omega_{n}\tau/\hbar}\ensuremath{\phi_{1}}\left({\mathbf k},\omega_{n}\right)},\\
\text{\ensuremath{\phi_{2}}\ensuremath{\left(\mathbf r,\tau\right)}} & =&\ensuremath{\sqrt{\frac{1}{\beta \hbar \mathcal{V}}}}\ensuremath{\sum_{{\mathbf k},\omega_{n}}e^{i{\mathbf k}\cdot {\mathbf r}-i\omega_{n}\tau/\hbar}\ensuremath{\phi_{2}}\left({\mathbf k},\omega_{n}\right)},
\end{eqnarray}
\end{subequations}
with $\beta$ defined as $1/k_{\text{B}}T$,
where $k_{\text{B}}$ is the Boltzmann constant and $T$ represents the temperature, 
$\mathcal{V}$ denotes the volume of the system under consideration.
Furthermore,
$k$ signifies the magnitude of the wave vector $\mathbf{k}$,
while $\omega_{n}=2\pi n/\beta$ represents the bosonic Matsubara frequency, with $n$ being integers.
This transformation facilitates a more streamlined analysis and allows us to leverage the properties of Fourier transforms in our calculations. 

\begin{figure}[H]
  \centering
  \includegraphics[scale=0.7]{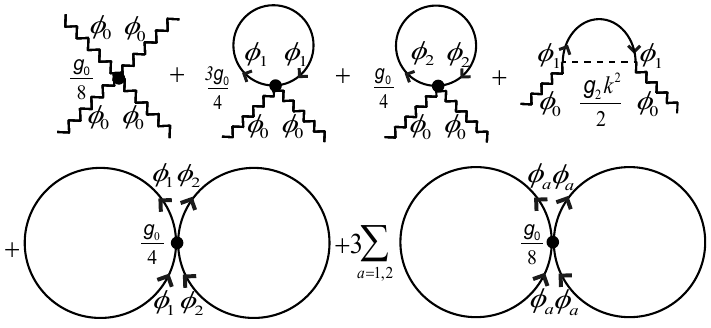}
  \caption{Feynman 2PI diagrams corresponding to Luttinger-Ward functional of $\Phi\left[\phi_0,G\right]$ in Eq.~(\ref{Veff}).}\label{b}
\end{figure}

Based on the CJT effective field theory, we find the CJT effective
potential from the Lagrangian density~(\ref{Lagrangian density}) as
\begin{eqnarray}
\label{Veff}
V_{\text{eff}}\left[\phi_{0},G\right]= & -&\frac{\mu}{2}\phi_{0}^{2}\nonumber \\
 & +&\frac{1}{2}\int_{\beta}\text{\text{Tr}}\left[\ln G^{-1}\left(k\right)+D_{0}^{-1}\left(k\right)G\left(k\right)-\mathbf{1}\right]\nonumber \\
 & +&\Phi\left[\phi_0,G\right],
\end{eqnarray}
with $\Phi\left[\phi_0,G\right]$ being the Luttinger-Ward functional as shown in Fig.~{\ref{b}}.
More specifically, $\Phi\left[\phi_0,G\right]=\frac{g_{0}}{8}\phi_{0}^{4}+\frac{3g_{0}}{4}\phi_{0}^{2}P_{11}+\frac{g_{0}}{4}\phi_{0}^{2}P_{22}+\frac{g_2}{2}\phi_{0}^{2}k^2P_{11}+\frac{g_{0}}{4}P_{11}P_{22}+\frac{3g_{0}}{8}\left(P_{11}^{2}+P_{22}^{2}\right)$,
where the functions $P_{11}$ and $P_{22}$ denote as
\begin{subequations}
\label{P11&P22}
\begin{eqnarray}
P_{11} & =&\int_{\beta}G_{11}\left(k\right)=\frac{1}{\beta}\sum_{n=-\infty}^{+\infty}\int\frac{d\mathbf{k}}{\left(2\pi\right)^{3}}G_{11}\left(k\right),\label{P11}\\
P_{22} & =&\int_{\beta}G_{22}\left(k\right)=\frac{1}{\beta}\sum_{n=-\infty}^{+\infty}\int\frac{d\mathbf{k}}{\left(2\pi\right)^{3}}G_{22}\left(k\right)\label{P22}.
\end{eqnarray}
\end{subequations}
Prior to utilizing the effective potential $V_\text{eff}$ 
for any computational purposes, 
it is imperative to engage in a brief discourse regarding its behavior under various loop approximation conditions. 
This discussion will provide valuable insights into the applicability and limitations of $V_\text{eff}$ 
within different theoretical frameworks.
\subsection{Zero-loop}\label{zeroloop}
In zero-loop approximation, we solely consider the first diagram depicted in Fig.~\ref{b} 
as the contribution to $\Phi\left[\phi_0\right]$. 
Consequently, the effective potential can be succinctly expressed as
$V_\text{eff}[\phi_0]=-\frac{\mu}{2}\phi_{0}^{2}+\frac{g_{0}}{8}\phi_{0}^{4}$.
Notably, $V_\text{eff}[\phi_0]$ involves a single variable $\phi_0$,
which represents the mean-field. By minimizing the $V_\text{eff}[\phi_0]$
with respect to $\phi_{0}$ and incorporating thermodynamic relationships, we derive
$\mu=g_{0}\rho$ and $\rho_{\text{ex}}=0$.
These results align perfectly with the well-established findings within the mean-field approximation framework.
\subsection{One-loop}\label{oneloop}
Subsequently, we proceed to truncate $\Phi\left[\phi_0,G\right]$ within one-loop approximation. 
This entails focusing solely on the diagrams featured in the first line of Fig.~\ref{b}. 
Under this approximation, the effective potential
$V_\text{eff}[\phi_0,G]$ takes on a specific form, which can be written as
\begin{eqnarray}
\label{Veff within one loop}
 V_\text{eff}[\phi_0,G]=&-&\frac{\mu}{2}\phi_{0}^{2}+\frac{g_{0}}{8}\phi_{0}^{4}\nonumber\\
&+&\frac{1}{2}\int_{\beta}\text{\text{Tr}}\left[\ln G^{-1}\left(k\right)+D_{0}^{-1}\left(k\right)G\left(k\right)-\mathbf{1}\right]\nonumber\\
&+&\frac{3g_{0}}{4}\phi_{0}^{2}P_{11}+\frac{g_{0}}{4}\phi_{0}^{2}P_{22}+\frac{g_2}{2}\phi_{0}^{2}k^2P_{11}.
\end{eqnarray}
In Equation~(\ref{Veff within one loop}), $D_{0}^{-1}\left(k\right)$ can be written as
\begin{equation}
 \label{D0,-1 in appendix}
D_{0}^{-1}\!\left(k\right)  =\begin{bmatrix}\frac{\hbar^{2}k^{2}}{2m}-\mu & -\omega_{n}\\
\omega_{n} & \frac{\hbar^{2}k^{2}}{2m}-\mu
\end{bmatrix},
\end{equation}
being the inversion propagator in free space. And $G\left(k\right)$ is the propagator of the system.
Notably, Equation~(\ref{Veff within one loop}) admits simplification 
by consolidating the terms in the third line into the second line. 
Through meticulous calculations, we can re-express $V_\text{eff}[\phi_0,G]$ in a more concise form as
\begin{eqnarray}
\label{Veff within one loop G_0^-1}
 V_\text{eff}[&\phi_0&,G]=-\frac{\mu}{2}\phi_{0}^{2}+\frac{g_{0}}{8}\phi_{0}^{4}\nonumber\\
&+&\frac{1}{2}\int_{\beta}\text{\text{Tr}}\left[\ln G^{-1}\left(k\right)+G_{0}^{-1}\left(k\right)G\left(k\right)-\mathbf{1}\right],
\end{eqnarray}
where $G_{0}^{-1}\left(k\right)$ is
\begin{equation}
 \label{G0,-1}
\!\!\!G_{0}^{-1}\left(k\right)  \!\!=\!\!\begin{bmatrix}\frac{\hbar^{2}k^{2}}{2m}\!\!-\!\!\mu\!+\!\!\frac{3g_{0}}{2}\phi_{0}^{2}\!\!+\!\!g_{2}\phi_{0}^{2}k^{2} & -\omega_{n}\\
\omega_{n} & \frac{\hbar^{2}k^{2}}{2m}\!\!-\!\!\mu\!+\!\!\frac{g_{0}}{2}\phi_{0}^{2}
\end{bmatrix}\!\!,
\end{equation}   
being the inversion propagator within one-loop approximation. Minimizing the effective potential $V_{\text{eff}}\left[\phi_{0},G\right]$
with respect to the elements of the propagator $G\left(k\right)$, we obtain that $G^{-1}\left(k\right)=G_{0}^{-1}\left(k\right)$.
Consequently, we reformulate $V_\text{eff}[\phi_0,G]$ as $V_\text{eff}[\phi_0,G_0]$, which in this approximation takes the form:
$V_\text{eff}[\phi_0,G_0]=-\frac{\mu}{2}\phi_{0}^{2}+\frac{g_{0}}{8}\phi_{0}^{4}
 +\frac{1}{2}\int_{\beta}\text{\text{Tr}}\left[\ln G^{-1}_{0}\left(k\right)\right]$. 
Further minimization of $V_\text{eff}[\phi_0,G_0]$ with respect to order parameter $\phi_{0}$ 
and application of thermodynamic relationships lead to the quantum depletion:
$\rho_{\text{ex}}=\frac{8\rho}{3\sqrt{\pi}}\left(\rho a_{\text{s}}^{3}\right)^{1/2}
-64\sqrt{\pi}\rho\frac{r_{\text{s}}}{a_{\text{s}}}\left(\rho a_{\text{s}}^{3}\right)^{3/2}$.
This result aligns with the findings reported in Ref.~\cite{Tononi2018}, 
thereby validating the one-loop approximation approach.
\subsection{Two-loop}\label{twoloop}
Upon substituting the expression of $\Phi\left[\phi_0,G\right]$ into the Eq.~(\ref{Veff}),
which encompasses all the diagrams depicted in Fig.~\ref{b}, we are able to recast $V_\text{eff}$ 
in an alternative form as
\begin{eqnarray}
\label{the orginal Veff}
V_{\text{eff}}[\phi_0,G]= & -&\frac{\mu}{2}\phi_{0}^{2}+\frac{g_{0}}{8}\phi_{0}^{4}\nonumber \\
 & +&\frac{1}{2}\int_{\beta}\text{\text{Tr}}\left[\ln G^{-1}\left(k\right)+G_{0}^{-1}\left(k\right)G\left(k\right)-\mathbf{1}\right]\nonumber \\
 & +&\frac{3g_{0}}{8}\left(P_{11}^{2}+P_{22}^{2}\right)+\frac{g_{0}}{4}P_{11}P_{22}.
\end{eqnarray}
In equation (\ref{the orginal Veff}), $G\left(k\right)$ is the propagator or Green's function. 
By minimizing the CJT effective potential $V_{\text{eff}}\left[\phi_{0},G\right]$
concerning the elements of the propagator $G\left(k\right)$, we obtain
\begin{equation}
\label{propergater G-1}
G^{-1}\left(k\right)=G_{0}^{-1}\left(k\right)+\Sigma,
\end{equation}
in which
\begin{equation}
\label{sum}
\Sigma=\begin{bmatrix}\Sigma_{1} & 0\\
0 & \Sigma_{2}
\end{bmatrix},
\end{equation}
with the matrix entries $\Sigma_{1}$ and $\Sigma_{2}$ being the self-energies
that can be constructed from Eqs.~(\ref{the orginal Veff}),~(\ref{P11}), and~(\ref{P22}),
\begin{subequations}
\begin{eqnarray}
\Sigma_{1}&=&\frac{3g_{0}}{2}P_{11}+\frac{g_{0}}{2}P_{22},\label{sum1}\\
\Sigma_{2}&=&\frac{3g_{0}}{2}P_{22}+\frac{g_{0}}{2}P_{11}.\label{sum2}
\end{eqnarray}
\end{subequations}
Furthermore, the expression for $G_{0}^{-1}\left(k\right)$ remains identical to that presented 
in Eq.~(\ref{G0,-1}) in Appendix~\ref{oneloop}. 
Consequently, by examining the poles of the Green's function, as detailed in Ref.~\cite{Andersen2004}, 
we can derive the dispersion relation, which is given as\begin{eqnarray}
\label{dispersion relation}
\epsilon_{\text{k}}&= & \left(\frac{\hbar^{2}k^{2}}{2m}-\mu+\frac{3g_{0}}{2}\phi_{0}^{2}+g_{2}\phi_{0}^{2}k^{2}+\Sigma_{1}\right)^{1/2}\nonumber\\
 &~& \times\left(\frac{\hbar^{2}k^{2}}{2m}-\mu+\frac{g_{0}}{2}\phi_{0}^{2}+\Sigma_{2}\right)^{1/2}.
\end{eqnarray}
By optimizing the CJT effective potential with respect to the order parameter
$\phi_{0}$, we can derive the gap equation within the Hartree-Fock (HF) approximation, 
\begin{equation}
\label{the Eq. of mu}
-\mu+\frac{g_{0}}{2}\phi_{0}^{2}+\Sigma_{1}=0.
\end{equation}

The Goldstone theorem \cite{Goldstone1962} postulates the necessity of a gapless excitation spectrum. 
Nevertheless, the dispersion relation derived from Eqs.~(\ref{dispersion relation}) and~(\ref{the Eq. of mu}) 
indicates a non-gapless spectrum, thereby violating the Goldstone theorem in the context of spontaneously broken symmetry 
within the HF approximation. 
To rectify this bug and reinstate the Nambu-Goldstone boson, we introduce a corrective term, denoted as $\Delta V$~\cite{Ivanov2005} 
to the $V_{\text{eff}}\left[\phi_{0},G\right]$, which is given by
\begin{align}
\label{delta V}
\Delta V & =-\frac{g_{0}}{4}\left(P_{11}^{2}+P_{22}^{2}\right)+\frac{g_{0}}{2}P_{11}P_{22}.
\end{align}
So that the revised $V_{\text{eff}}$ is
\begin{eqnarray}
\label{the revised Veff}
V_{\text{eff}}= & -&\frac{\mu}{2}\phi_{0}^{2}+\frac{g_{0}}{8}\phi_{0}^{4}\nonumber \\
 & +&\frac{1}{2}\int_{\beta}\text{\text{Tr}}\left[\ln G^{-1}\left(k\right)+G_{0}^{-1}\left(k\right)G\left(k\right)-\mathbf{1}\right]\nonumber \\
 & +&\frac{g_{0}}{8}\left(P_{11}^{2}+P_{22}^{2}\right)+\frac{3g_{0}}{4}P_{11}P_{22}.
\end{eqnarray}
In Equation~(\ref{the revised Veff}), the first two terms constitute the mean-field component, 
corresponding to the condensate atoms, while the subsequent terms represent the excitation part stemming from the excited atoms. 
By replicating the calculations performed leading up to Eq.~(\ref{propergater G-1}), we obtain the revised inverse propagator,
\begin{equation}
\label{the revised G-1}
G^{-1}\left(k\right)=G_{0}^{-1}\left(k\right)+\Sigma^{\prime},
\end{equation}
where
\begin{equation}
\label{sum prime}
\Sigma^{\prime}=\begin{bmatrix}\Sigma_{1}^{\prime} & 0\\
0 & \Sigma_{2}^{\prime}
\end{bmatrix},
\end{equation}
in which the self-energies $\Sigma_{1}^{\prime}$ and $\Sigma_{2}^{\prime}$ are defined as follows,
\begin{subequations}
\begin{eqnarray}
\Sigma_{1}^{\prime} & =&\frac{g_{0}}{2}P_{11}+\frac{3g_{0}}{2}P_{22}\label{sum1 prime},\\
\Sigma_{2}^{\prime} & =&\frac{g_{0}}{2}P_{22}+\frac{3g_{0}}{2}P_{11}\label{sum2 prime}.
\end{eqnarray}
\end{subequations}
Then, the gap equation that $\mu$ satisfies can be reformulated as
\begin{equation}
\label{rewrite the Eq. of mu}
-\mu+\frac{g_{0}}{2}\phi_{0}^{2}+\Sigma_{2}^{\prime}=0.
\end{equation}
Furthermore, we can formulate the Schwinger-Dyson (SD) equation as
\begin{equation}
\label{Dyson equation}
-\mu+\frac{3g_{0}}{2}\phi_{0}^{2}+\Sigma_{1}^{\prime}=M^{2}.
\end{equation}
Upon modifying $m$ to $m^{*}=m/\left(1+2mg_{2}\phi_{0}^{2}/\hbar^{2}\right)$, 
the revised inversion propagator $G^{-1}\left(k\right)$ and the corresponding propagator $G\left(k\right)$ can be turned out
\begin{subequations}
\label{G-1&G}
\begin{eqnarray}
\label{the revised G-1,k}
G^{-1}\left(k\right)=\begin{bmatrix}\frac{\hbar^{2}k^{2}}{2m^{*}}+M^{2}, & -\omega_{n}\\
\omega_{n}, & \frac{k^{2}}{2m}
\end{bmatrix},\\
 \label{the revised G,k}
G\left(k\right)=\frac{1}{\epsilon_{\text{k}}^{2}+\omega_{n}^{2}}\begin{bmatrix}\frac{\hbar^{2}k^{2}}{2m}, & \omega_{n}\\
-\omega_{n}, & \frac{\hbar^{2}k^{2}}{2m^{*}}+M^{2}
\end{bmatrix}.   
\end{eqnarray}
\end{subequations}
In Equation~(\ref{the revised G,k}), $\epsilon_{\text{k}}=\sqrt{\frac{\hbar^{2}k^{2}}{2m}\left(\frac{\hbar^{2}k^{2}}{2m^{*}}+M^{2}\right)}$, 
representing the dispersion relation. 

Plugging Eqs.~(\ref{G0,-1}) and~(\ref {the revised G,k}) into Eq.~(\ref{the revised Veff}) and 
performing the calculations pertaining to the second term,
we proceed to obtain
\begin{eqnarray}
\label{the final Veff in appendix}
V_{\text{eff}}= & -&\frac{\mu}{2}\phi_{0}^{2}+\frac{g_{0}}{8}\phi_{0}^{2}\nonumber \\
 & +&\frac{1}{2}\int_{\beta}\text{\text{Tr}}\left[\ln G^{-1}\left(k\right)\right]\nonumber \\
 & +&\frac{g_{0}}{8}\left(P_{11}^{2}+P_{22}^{2}\right)+\frac{3g_{0}}{4}P_{11}P_{22}\nonumber \\
 & +&\frac{1}{2}\left(-\mu+\frac{3g_{0}}{2}\phi_{0}^{2}-M^{2}\right)P_{11}\nonumber \\
 & +&\frac{1}{2}\left(-\mu+\frac{g_{0}}{2}\phi_{0}^{2}\right)P_{22}.
\end{eqnarray}
Meanwhile, by evaluating Eq.~(\ref{P11&P22}) through summing over the bosonic Matsubara 
frequencies using Eq.~(\ref{the revised G,k}), 
and then taking the limit as $T\rightarrow 0$,
we perform the calculations in spherical coordinates employing dimensional regularization techniques, yielding:
\begin{subequations}
\label{P11&P22 about M}
\begin{eqnarray}
\label{P11 about M}
P_{11} &=&\frac{1}{2}\int\frac{d\mathbf{k}}{\left(2\pi\right)^{3}}\sqrt{\frac{\frac{\hbar^{2}k^{2}}{2m}}{\frac{\hbar^{2}k^{2}}{2m^{*}}+M^{2}}}\nonumber\\
&=&\frac{1}{2}\int\frac{d\mathbf{k}}{\left(2\pi\right)^{3}}\sqrt{\frac{\frac{\hbar^{2}k^{2}}{2m}}{\frac{\hbar^{2}k^{2}}{2m^{*}}+M^{2}}}\sqrt{\frac{m^*}{m}}\nonumber\\
&=&\frac{1}{8\pi^{2}}\left(\frac{2m^{*}M^{2}}{\hbar^{2}}\right)^{\frac{3}{2}}\sqrt{\frac{m^*}{m}}\int_{0}^{\infty}t\left(t+1\right)^{-\frac{1}{2}}dt\nonumber\\
&=&\frac{1}{8\pi^{2}}\left(\frac{2m^{*}M^{2}}{\hbar^{2}}\right)^{\frac{3}{2}}\sqrt{\frac{m^*}{m}}\frac{\Gamma\left[2\right]\Gamma\left[\frac{-3}{2}\right]}{\Gamma\left[\frac{1}{2}\right]}\nonumber\\
&=&\frac{\sqrt{2}m^{*3/2}M^{3}}{3\pi^{2}\hbar^{3}}\sqrt{\frac{m^*}{m}},\\
\label{P22 about M}
P_{22}
&=&\frac{1}{2}\int\frac{d\mathbf{k}}{\left(2\pi\right)^{3}}\sqrt{\frac{\frac{\hbar^{2}k^{2}}{2m^{*}}+M^{2}}{\frac{\hbar^{2}k^{2}}{2m}}}\nonumber\\
&=&\frac{1}{2}\int\frac{d\mathbf{k}}{\left(2\pi\right)^{3}}\sqrt{\frac{\frac{\hbar^{2}k^{2}}{2m^{*}}+M^{2}}{\frac{\hbar^{2}k^{2}}{2m^{*}}}}\sqrt{\frac{m}{m^*}}\nonumber\\
&=&\frac{1}{8\pi^{2}}\left(\frac{2m^{*}M^{2}}{\hbar^{2}}\right)^{\frac{3}{2}}\sqrt{\frac{m}{m^*}}\int_{0}^{\infty}t^{0}\left(t+1\right)^{\frac{1}{2}}dt\nonumber\\
&=&\frac{1}{8\pi^{2}}\left(\frac{2m^{*}M^{2}}{\hbar^{2}}\right)^{\frac{3}{2}}\sqrt{\frac{m}{m^*}}\frac{\Gamma\left[1\right]\Gamma\left[\frac{-3}{2}\right]}{\Gamma\left[-\frac{1}{2}\right]}\nonumber\\
&=&-\frac{m^{*3/2}M^{3}}{3\sqrt{2}\pi^{2}\hbar^{3}}\sqrt{\frac{m}{m^*}}.
\end{eqnarray}
\end{subequations}
\section{DETAILED CALCULATION OF THE SOLUTION OF EQ.~(\ref{the solution of $M$})}\label{DCTS}
For the sake of self-consistence of this work, we briefly review the key steps to solve the Eq.~(\ref{the solution of $M$}) 
utilizing perturbation theory. 
Initially, we introduce a small parameter $\epsilon$ to the left-hand side of $M^3$, thereby transforming the cubic equation in
$M$ into the form
\begin{eqnarray}\label{the M of epsilon}
\epsilon M^{3}&+&\frac{3\pi\hbar\left(1+4\frac{mg_{2}}{\hbar^{2}}\rho\right)^{3/2}}{8\sqrt{2}a_{\text{s}}m^{*1/2}}M^{2}\nonumber\\
&-&\frac{3\pi^{2}\hbar^{3}\left(1+4\frac{mg_{2}}{\hbar^{2}}\rho\right)^{1/2}}{\sqrt{2}m^{*3/2}}\rho=0.
\end{eqnarray}
Meanwhile, we express $M$ as a polynomial expansion in terms of the parameter $\epsilon$
\begin{equation}\label{expansion of M}
M\rightarrow M_{0}+\epsilon M_{1}+\epsilon^{2}M_{2}.
\end{equation}

Next, we substitute the expression for $M$ from Eq.~(\ref{expansion of M}) into Eq.~(\ref{the M of epsilon}) 
and organize the resulting equation into orders of $\epsilon^{0}$, $\epsilon^{1}$ and $\epsilon^{2}$.

(i) $\epsilon^{0}$ order of equation
 \begin{equation}\label{epsilon0 order}
   \frac{3\pi\hbar\left(1+4\frac{mg_{2}}{\hbar^{2}}\rho\right)^{3/2}}{8\sqrt{2}a_{\text{s}}m^{*1/2}}M_{0}^{2}-\frac{3\pi^{2}\hbar^{3}\left(1+4\frac{mg_{2}}{\hbar^{2}}\rho\right)^{1/2}}{\sqrt{2}m^{*3/2}}\rho=0,
 \end{equation}
 \begin{equation}\label{M0}
  \Rightarrow M_{0}=\sqrt{2g_{0}\rho }.
 \end{equation}
 
 (ii) $\epsilon^{1}$ order of equation
\begin{equation}\label{epsilon1 order}
  M_{0}^{3}+\frac{3\pi\hbar\left(1+4\frac{mg_{2}}{\hbar^{2}}\rho\right)^{3/2}}{4\sqrt{2}a_{\text{s}}m^{*1/2}}M_{0}M_{1}=0,
 \end{equation}
 \begin{equation}\label{M1}
  \Rightarrow M_{1}=-\frac{16\sqrt{2g_{0}\rho }}{3\sqrt{\pi}\left(1+4\frac{mg_{2}}{\hbar^{2}}\rho\right)^{2}}\sqrt{\rho a_{\text{s}}^{3}}.
 \end{equation}
 
 (iii) $\epsilon^{2}$ order of equation
\begin{equation}\label{epsilon2 order}
 3M_{0}^{2}M_{1}+\frac{3\pi\hbar\left(1+4\frac{mg_{2}}{\hbar^{2}}\rho\right)^{3/2}}{8\sqrt{2}a_{\text{s}}m^{*1/2}}\left(M_{1}^{2}+2M_{0}M_{2}\right)=0,
 \end{equation}
 \begin{equation}\label{M2}
 \Rightarrow M_{2}=\frac{640\sqrt{2g_{0}\rho}}{9\pi\left(1+4\frac{mg_{2}}{\hbar^{2}}\rho\right)^{4}}\rho a_{\text{s}}^{3}.
 \end{equation}
 
Finally, we derive the expression for $M$, which is expanded as $M_0+M_1+M_2$, in terms of the gas parameter $\rho a_{\text{s}}^{3}$
 \begin{widetext}
\begin{equation}
\label{the solution of $M$ in appendix}
M=\sqrt{2g_{0}\rho }\left\{1-\frac{16}{3\sqrt{\pi}\left(1+4\frac{mg_{2}}{\hbar^{2}}\rho\right)^{2}}\sqrt{\rho a_{\text{s}}^{3}}\left[1-\frac{40}{3\sqrt{\pi}\left(1+4\frac{mg_{2}}{\hbar^{2}}\rho\right)^{2}}\sqrt{\rho a_{\text{s}}^{3}}\right]+\mathcal{O}\left[\left(\rho a_{\text{s}}^{3}\right)^{3/2}\right]\right\}.
\end{equation}
\end{widetext}
\bibliography{loopy}

\end{document}